\documentclass[10pt, doublecolumn]{IEEEtran}
\usepackage{float}
\usepackage{graphicx}
\usepackage{stfloats}
\usepackage{caption}
\usepackage[ruled]{algorithm2e}
\ifCLASSOPTIONcompsoc
  \usepackage[caption=false,font=normalsize,labelfont=sf,textfont=sf]{subfig}
\else
  \usepackage[caption=false,font=footnotesize]{subfig}
\fi
\usepackage{indentfirst}
\usepackage{graphicx}
\usepackage{bm}
\usepackage{amsmath}
\allowdisplaybreaks[4]
\usepackage{amssymb}
\usepackage{times}
\usepackage{mathtools}
\usepackage{psfrag}
\usepackage{cite}
\usepackage{lastpage}
\usepackage{fancyhdr}
\usepackage{color}
 \usepackage{amsthm}
\usepackage{bigints}
\newtheorem{Theorem}{Theorem}
\newtheorem{Lemma}{Lemma}
\theoremstyle{remark}
\newtheorem{Remark}{$\quad$Remark}

\begin{document}
\title{Age of Information Analysis for CR-NOMA Aided Uplink Systems with Randomly Arrived Packets}
\author{Yanshi Sun, \IEEEmembership{Member, IEEE}, Yanglin Ye, Zhiguo Ding, \IEEEmembership{Fellow, IEEE}, Momiao Zhou, \IEEEmembership{Member, IEEE}, Lei Liu
\thanks{
The work of Y. Sun and Y. Ye was supported in part by Hefei University of Technology's
construction funds for the introduction of talents with funding number 13020-03712022011
and the National Natural Science Foundation of China under Grant 62301208.
The work of L. Liu was supported in part by the National Natural Science Foundation of China
under Grants 62201456.
This article was presented in part at
the 2024 IEEE Wireless Communications and Networking Conference (WCNC)
\cite{sun2024age}.

Y. Sun, Y. Ye and M. Zhou  are  with the School of Computer Science and Information
Engineering, Hefei University of Technology, Hefei, 230009, China. (email: sys@hfut.edu.cn, yanglinyeuna@163.com and mmzhou@hfut.edu.cn).

Z. Ding is with Department of Computer and Information Engineering,
Khalifa University, Abu Dhabi, UAE. (email: zhiguo.ding@ieee.org).

L. Liu is with the Shaanxi Key Laboratory of
Information Communication Network and Security, Xi'an University of Posts
and Telecommunications, Xi'an 710121, China (e-mail: leiliu@xupt.edu.cn).
}\vspace{-0.8em}}
\maketitle
\begin{abstract}
This paper studies the application of  cognitive radio inspired non-orthogonal multiple access (CR-NOMA) to reduce age of information (AoI) for uplink transmission.
In particular, a time division multiple access (TDMA) based legacy network is considered, where each user is allocated with a dedicated time slot to transmit its status update information. The CR-NOMA is implemented as an add-on to the TDMA legacy network, which enables each user to have more opportunities to transmit by sharing other user's time slots. A rigorous analytical framework is developed to obtain the expressions for AoIs achieved by CR-NOMA with and without re-transmission, by taking the randomness of the status update generating process into consideration. Numerical results are presented to verify the accuracy of the developed analysis. It is shown that the AoI can be significantly reduced by applying CR-NOMA compared to TDMA.
Moreover, the use of re-transmission is helpful to reduce AoI, 
especially when the status arrival rate is low.
\end{abstract}

\section{Introduction}
With a rapid development of wireless communications, establishing ubiquitous connectivity for massive machine type communications (mMTC) becomes feasible \cite{you2021towards,abbas2020joint}. Under this background, more and more real-time applications for monitoring and controlling are emerging, e.g., autonomous driving. In these scenarios,
information sources (such as sensors) need to frequently transmit their status updates to destinations,
in order to keep the status information collected by the destinations as freshness as possible. Because the fresher the status information is, the more conducive it is for making correct decisions. To this end, the concept of age of information (AoI) has been recently proposed as a new metric to characterize the timeliness of status updating systems \cite{kaul2012real}. In particular, AoI is defined as the time duration of the newest status update observed at the receiver since its generation. Existing literature shows that minimizing AoI is not equivalent to maximizing utilization (throughput) or minimizing status packet delivery delay. Due to the above reasons,
the study of AoI has raised considerable attention from both academia and industry \cite{kaul2012real,sun2017update,yates2021age,abbas2021markovian}.

The AoI in single-source scenarios has been extensively investigated in the literature \cite{costa2016age,inoue2019general,cao2021can,zheng2021decode}. However, in multi-source scenarios, due to the limited degrees of freedom (DoF) for wireless transmission, the devices need to share the channel resource blocks to complete their status updating transmissions. As a result, the AoI achievable for a certain source depends heavily on the adopted multiple access (MA) technique, which determines how channel resource blocks are allocated to multiple users \cite{yates2018age,chen2022age,yates2017status}. Orthogonal multiple access (OMA) is a very straightforward way to avoid inter-user interferences, and has been widely used in communication networks.
In \cite{pan2020information}, the achievable AoIs for time division multiple access (TDMA) and frequency division multiple access (FDMA) were investigated, which shows that TDMA outperforms FDMA in terms of average AoI, while
FDMA is better in terms of stability under time-varying channels.

Different from OMA, non-orthogonal multiple access (NOMA) allows multiple users to transmit signals simultaneously by occupying the same channel resource block. It is shown by the literature that, compared to OMA, NOMA is more spectral efficient, and more supportive for massive connectivity and low latency \cite{saito2013non,ding2017survey, ding2018impact}. Therefore, it is important to investigate the role of NOMA to reduce AoI in status updating systems \cite{maatouk2019minimizing, abbas2022minimizing, liu2021peak,feng2022optimizing,gao2022non,zhang2021age}.
{\color{black}In \cite{maatouk2019minimizing} and \cite{abbas2022minimizing}, dynamic policies to switch between NOMA and OMA were developed to minimize AoI.
In \cite{liu2021peak}, the achievable peak AoI for NOMA with the first-come-first-serve (FCFS) queuing rule was studied, where sources are ordered according to their distances to the base station.
In \cite{feng2022optimizing}, AoI was optimized for a reconfigurable intelligent surface
 (RIS) assisted NOMA network by using tools from reinforcement learning.
In \cite{gao2022non}, NOMA based AoI in low earth orbit (LEO) satellite-terrestrial integrated networks was
investigated, where average AoI minimization in terrestrial
 networks and  average AoI minimization among satellites were both considered.

In \cite{ding2023age}, cognitive radio inspired NOMA (CR-NOMA), as a very important form of NOMA, has also been applied to reduce AoI in status updating systems.
The key idea of CR-NOMA is that one user have additional transmission opportunities as a secondary user by sharing other users' resource blocks. Compared to existing NOMA schemes considered in \cite{maatouk2019minimizing, abbas2022minimizing, liu2021peak,feng2022optimizing,gao2022non}, a very appealing feature of CR-NOMA is
its simplicity of implementation. Specifically,  CR-NOMA can be implemented as a simple add-on to a legacy network based on OMA, with very limited modifications to the legacy network.
It has been shown by \cite{ding2023age} and \cite{ding2023new}, CR-NOMA can play an important role to reduce AoI.}
Particularly, in \cite{ding2023age}, CR-NOMA was implemented over a time division multiple access (TDMA) based legacy network, where each user is allocated with a single dedicated time slot in each frame, and each user is offered one additional opportunity to transmit within each frame by sharing its partner's time slot. Two data generation models were considered, namely generate-at-will (GAW) and generate-at-request (GAR), where
the GAW model assumes that a new status update is generated right before each transmit time slot, and the GAW
model assumes that a new status update is generated right at the beginning of each frame.
However, GAW and GAR are ideal models, which might be unrealistic in many practical scenarios where the status data is generated randomly. To the author's best knowledge, how to characterize the AoI performance of the CR-NOMA
assisted status updating system with random arrivals is still open, which motivates this paper.

This paper aims to investigate the average AoI achievable for the CR-NOMA assisted status updating system when status arrives randomly. Similar to \cite{ding2023age}, a TDMA based legacy network is considered, based on which CR-NOMA
is carried out as an add-on. The main contributions of this paper are listed as follows.
\begin{itemize}
\item Different from the existing work \cite{ding2023age} which adopts an ideal data generation model, this paper
considers a more general model by capturing the randomness of the data generation. As a result, the queuing process of the waiting status data packets has to be considered, which is a new challenging problem compared to
the GAW and GAR models. Since only the newest status data affects the AoI at the receiver, this paper considers the
commonly used last-come-first-serve (LCFS) queuing rule. Besides, the strategies with and without retransmission are also considered in the paper, which is also a new challenging problem compared to the GAW and GAR models.
\item Through rigorous derivation, closed-form expressions for the average AoIs achieved by CR-NOMA with and
without re-transmission (termed ``NOMA-NRT'' and ``NOMA-RT'') are obtained. Besides, for the comparison purpose,  analyses for TDMA based schemes are also provided.  Note that, compared to the analyses for the GAW and GAR models,
the analysis for random arrival model is much more challenging, especially for NOMA-RT, due to the fact that
each user's queuing buffer state and data transmission reliability are coupled with those of its partner.
\item Simulation results are provided to validate the developed analytical results. Comparisons of the considered CR-NOMA schemes with the existing TDMA based schemes are also provided. It is shown that the achievable average AoI can be significantly reduced by applying CR-NOMA. Furthermore, retransmission is necessary to reduce AoI, especially when the data arrival rate is low. Moreover, the impact of system parameters on AoI, such as the data arrival rate and the duration of a time slot, has also been demonstrated and discussed.
\end{itemize}

The remainder of this paper is organized as follows. In Section II, the system model and the considered transmission strategy are described. In Section III, analytical frameworks are developed to characterize the average AoI achieved by the considered transmission strategies. Simulation results are presented in Section IV. Finally, the paper is concluded in Section V.
\section{System model}
\subsection{Update arrival and queuing process}
Consider a wireless communication scenario, where $M$ sources send their status updates to one receiver and each source is denoted by $U_m$, $1\leq m \leq M$. Random arrivals are considered for the status generation process. Specifically, status updates arrive at source $U_m$
according to a one-dimensional Poisson process with parameter $\lambda_m$ \footnote{{\color{black}Although this paper focuses on Poisson process for modeling the random arrivals, it is noteworthy that the developed analytical framework is also applicable to other arrival models, such as the Bernoulli model \cite{Asvadi2023} and GAW model \cite{ding2023age}}}.
It is assumed that each status update packet contains $N$ bits.
The channel resources are divided into consecutive time slots and are allocated to the sources.
The considered time slot allocation rules will be discussed later. Note that each source can only transmit its status update through the assigned transmitting time slots. Last-come-first-served (LCFS) queuing  is considered at each source.
Specifically, each source maintains a buffer with size one to save the latest update to be transmitted. If a new status update arrives at a source, the source will put the new update into the buffer by dropping the previously saved update information. At the beginning of each transmitting time slot, each source moves the status information saved in its queuing buffer to its transmitter, meanwhile the queuing buffer is set to be empty to accommodate future updates. It is noteworthy that if a new update comes during the transmitting time slot, it can be pushed into the queuing buffer, but it does not affect the transmission of the transmitted data.

\subsection{Multiple Access Strategies}
\subsubsection{TDMA}
This paper considers TDMA as the benchmark multiple access strategy. Specifically, the timeline is divided into consecutive time frames. In each TDMA time frame, each source is allocated a single time slot with duration $T$.
Without loss of generality, the $m$-th time slot in each frame is allocated to $U_m$.
Each source is allowed to transmit update information to the receiver only within the assigned time slot, if it has an
update data to transmit. Therefore, the achievable data rate of $U_m$ in  the $m$-th time slot of frame $i$ is given by:
\begin{align}
 R_{i,m}^m=\log\left(1+P|h_{i,m}^{m}|^2\right),
\end{align}
where $P$ is the transmit power, $h_{i,j}^{m}$ denotes the channel of $U_m$ in the $j$-th time slot of frame $i$. Note that, without loss of generality, the noise power is normalized in this paper.

\subsubsection{CR-NOMA}
CR-NOMA can be used as an add-on to TDMA to improve the freshness of the data collected at the receiver.
Particularly, in CR-NOMA, $U_m$ and $U_{m'}$ are paired together to form a NOMA group, where $1\leq m\leq \frac{M}{2}$ and $m'=m+\frac{M}{2}$. In each NOMA group, the paired users can share the channel resource block with each other.

Specifically, in the $m$-th time slot of frame $i$, $U_m$ and $U_{m'}$ are treated as the primary user and secondary user, respectively. Note that, $U_m$ transmits its signal with power $P$ as in TDMA, if it has update information to transmit.
Meanwhile, $U_{m'}$ can also transmit signal within the time slot by applying NOMA, if it has the updated information to transmit. The application of NOMA is transparent to the primary user. To this end, the secondary user's signal is decoded at the first stage of SIC \footnote{{\color{black}Although fixed SIC order is adopted in this paper, it is worth pointing out that advanced SIC methods, such
as hybrid SIC is helpful to further reduce AoI for the considered system \cite{ding2021new,sun2021new,sun2023hybrid}.}}, which can ensure that the primary user achieves the same transmission reliability as in TDMA. Hence, the achievable data rate of $U_{m'}$ in the $m$-th time slot of frame $i$ is given by:
\begin{align}
 R_{i,m}^{m'}=\log\left(1+\frac{P^s|h_{i,m}^{m'}|^2}{\delta_{i,m}^{m}P|h_{i,m}^{m}|^2+1}\right),
\end{align}
where $P^s$ is the transmit power of the secondary user, and $\delta_{i,j}^k$ is an indicator  variable to
denote whether source $m$ transmits a signal in the $j$-th time slot of frame $i$.

Similarly, in the $m'$-th time slot of frame $i$, $U_{m'}$ is treated as the primary user and $U_m$ is the secondary user. Following the same transmission strategy aforementioned above, $U_{m'}$ achieves the same transmission performance as in TDMA, and $U_m$ transmits signal opportunistically, yielding the following achievable data rate:
\begin{align}
 R_{i,m'}^{m}=\log\left(1+\frac{P^s|h_{i,m'}^{m}|^2}{\delta_{i,m'}^{m'}P|h_{i,m'}^{m'}|^2+1}\right).
\end{align}
One appealing feature of the considered CR-NOMA is explained as follows.
For the considered CR-NOMA scheme, when $U_m$ transmits in the $m$-th slot, its transmission success probability is given by:
\begin{align}\label{P_{mm}}
P_{mm}=&P\left(R_{i,m}^m > {N}/{T}\right)=e^{-\frac{\epsilon}{P}},
\end{align}
where $\epsilon=2^{N/T}-1$, which is the same as the transmission success probability of $U_m$ in the TDMA mode.
Similarly, for the considered CR-NOMA scheme, when $U_{m'}$ transmits in the $m'$-th slot, its transmission success probability is given by:
\begin{align}
P_{m'm'}=&P\left(R_{i,m'}^{m'} > {N}/{T}\right)=e^{-\frac{\epsilon}{P}},
\end{align}
which is also the same as the transmission success probability of $U_m$ if TDMA is adopted.

{\color{black}Besides, the application of CR-NOMA can also help to improve the bandwidth utilization.
The reasons are as follows. In TDMA schemes, each source is allocated with
a dedicated slot. If the source has no packet to transmit in a slot, then this slot cannot be utilized, resulting in a low utilization. However, when  the considered CR-NOMA schemes are applied,
if source $m$ has no packet to transmit in the $m$-th slot, the $m$-th slot can still be possibly utilized by
source $m'$. Thus, the bandwidth utilization can be improved by the considered CR-NOMA compared to TDMA.}

\begin{Remark}
{\color{black}In this paper, perfect channel state information (CSI) and ideal SIC are assumed for simplifying the analysis. However, it is noteworthy that the developed analytical framework can be extended to the cases where imperfect CSI and SIC are considered \cite{li2019residual}. Besides, to facilitate CSI estimation, it is necessary to set a mini-slot before the each user's transmission slot for pilot transmission, which may increase the AoI. However, due to reason that the length of the mini-slot is usually much shorter than that of the user's transmission slot, the impact of the CSI acquisition on AoI can be ignored. }
\end{Remark}

\subsection{With and without re-transmission}
\subsubsection{Without re-transmission}at the end of each transmitting slot, the transmitted data will be discarded by the transmitter, regardless of whether the data transmission is successful or not.
\subsubsection{With re-transmission}at the end of each transmitting slot, if the signal is not successfully transmitted and there is no new status update data arrives, the transmitted data will be moved back into the queuing buffer for re-transmission. Otherwise, the transmitted data will be discarded. Note that, if a newer status update comes before the next transmitting time slot, the previously received update data will be discarded, in order to improve the freshness of the data collected at the source \footnote{{\color{black}Please note that the utilization of retransmissions may result in higher energy consumption. In this paper, it is assumed that the additional energy consumption is affordable for the source, where the timeliness of the information is more important.}}.

For notational convenience, the TDMA scheme with and without re-transmission is termed ``TDMA-RT'' and ``TDMA-NRT'', respectively. And the NOMA scheme with and without re-retransmission is termed ``NOMA-RT'' and ``NOMA-NRT'', respectively.

\begin{Remark}
{\color{black}Some modifications to the legacy TDMA network is necessary for the implementation of CR-NOMA. First, if re-transmission strategy is adopted, the receiver needs to carry out a one-bit feedback to the users at the end each slot, to inform the users whether the re-transmission is needed. Besides, to ensure that the transmission of the secondary user to the corresponding primary user is transparent, the receiver needs to feedback the permitted maximal
data rate shown in (2) (or (3)) to the secondary user.}
\end{Remark}

\subsection{Performance metric}
\begin{figure}[!t]
  \centering
      \setlength{\abovecaptionskip}{0em}   
\setlength{\belowcaptionskip}{-2em}   
  \includegraphics[width=2.8in]{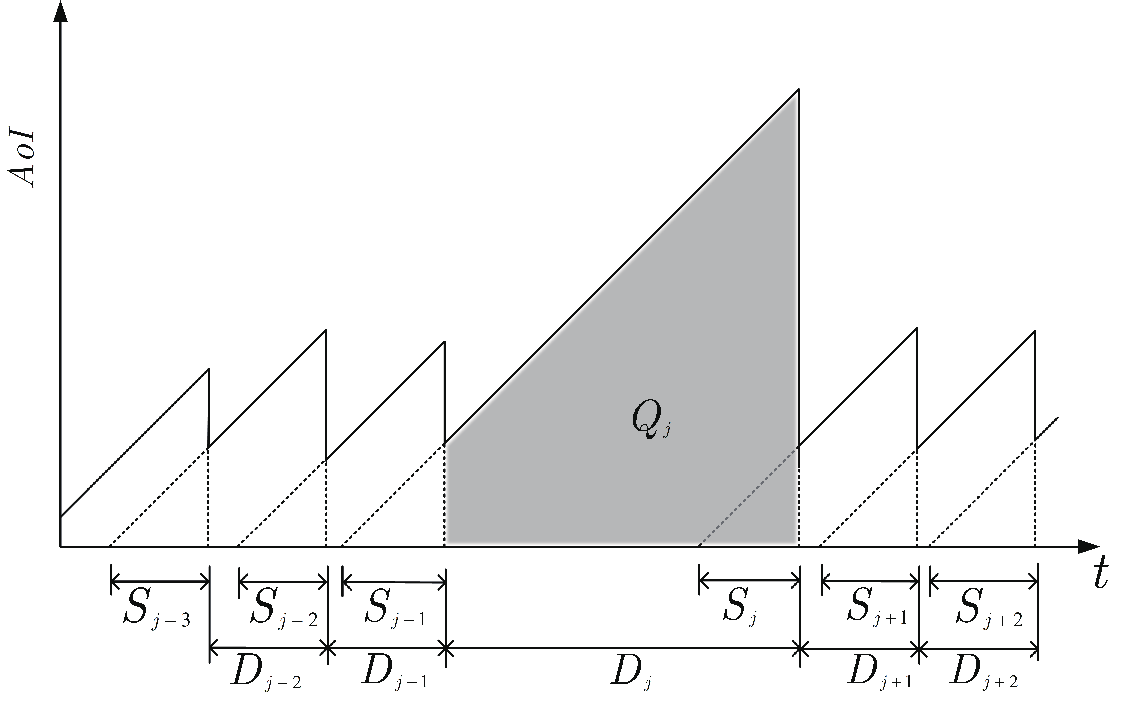}\\
  \caption{Illustration of the AoI of a status updating process.}\label{serration_exp}
\end{figure}
In this paper, the age of information (AoI) is used as the performance metric to evaluate the freshness of the
latest update which has been  successfully delivered to the receiver. Note that, only the status updates which are successfully delivered to the receiver affect the AoI. For a specific source $U_m$, the generation time of the
$j$-th successfully delivered status update packet is denoted by $t_j$, and its corresponding arrival time
at the receiver is denoted by $t_j'$.
The instantaneous AoI of source $U_m$'s update at the receiver is a time varying function, which is denoted by $\Delta_{m}(t)$ and  determined by the time difference between the current time and the generation time of the newest status update information observed at the receiver. Let $\Omega(t)=\max\{t_j|t_j'\leq t\}$ denote the index of the newest update observed at the receiver, then the instantaneous AoI of $U_m$ can be expressed as:
\begin{align}
 \Delta_{m}(t)=t-u(t),
\end{align}
where $u(t)=t_{\Omega(t)}$ is the generation time of the newest status update information observed at the receiver.
Note that, the age process $\Delta_m (t)$ forms a sawtooth path as illustrated in Fig. \ref{serration_exp}.

The average AoI of $U_m$ is defined as the average of AoI over time, which can be expressed as {\color{black}\cite{sun2017update}}:
\begin{align}\label{AAoI}
 \bar{\Delta}_{m}=\lim_{\mathcal{T}\rightarrow \infty} \frac{1}{\mathcal{T}}\int_{0}^{\mathcal{T}}\Delta_{m}(t)\,dt.
\end{align}
The evaluation of  $\bar{\Delta}_{m}$ can be described as follows.
For the ease of exposition, denote $D_j=t_j'-t_{j-1}'$ by the interval between the $j$-th and ($j-1$)-th
successful delivery, and  $S_j=t_j'-t_j$ by the system time of a successfully delivered update.
It can be straightforwardly verified that the evaluation of the average AoI is equivalent to find the sum of a
series of trapezoidal areas, denoted by $Q_j$ {\color{black}\cite{yates2021age}.} As shown in Fig. \ref{serration_exp}, where
\begin{align}\label{Q_j}
Q_j=D_jS_{j-1}+\frac{1}{2}(D_j)^2.
\end{align}
Then, the average AoI $\bar{\Delta}_m$ can be expressed as \cite{costa2016age}:
\begin{align}
\bar{\Delta}_m=\lim_{J\rightarrow \infty}\frac{\sum_{j=1}^{j=J}{Q_j}}{\sum_{j=1}^{j=J}{D_j}}.
\end{align}

Further, when $(D_j,S_j)$ is a stationary and ergodic process, the evaluation of $\bar{\Delta}_m$ can be simplified as:
\begin{align}\label{}
\bar{\Delta}_m=\frac{\mathbb{E}\{Q_j\}}{\mathbb{E}\{D_j\}}=\frac{\mathbb{E}\{D_jS_{j-1}\}+\mathbb{E}\{{D_j}^2\}/2}{\mathbb{E}\{D_j\}}.
\end{align}

\begin{Remark}
{\color{black}Note that discrete AoI metrics have been widely adopted in the literature for simplification.
However, such metrics are not applicable for this paper. Because random packet arrival model is considered in this paper, which means that the new status update packet may arrives at any instant of a frame. Thus, it is necessary for
the AoI metric to have the capability to quantify fractional duration of a time slot, which excludes the utilization
of discrete AoI metrics.}
\end{Remark}

\section{Analysis on AoI for TDMA-NRT, NOMA-NRT, TDMA-RT and NOMA-RT}
In this section, the average AoIs achieved by the TDMA-NRT, NOMA-NRT, TDMA-RT and NOMA-RT schemes are analyzed, respectively. Due to the symmetry among users, it is sufficient to focus on a particular user, say user $m$.
Compared to the schemes with retransmission, the analyses for TDMA-NRT and NOMA-NRT are relatively easier, since the whole time line can be split into consecutive and independent parts.
\begin{figure*}[!t]
\vspace{-2em}
  \centering
      \setlength{\abovecaptionskip}{0em}   
	\setlength{\belowcaptionskip}{-1em}
  \includegraphics[width=6.0in]{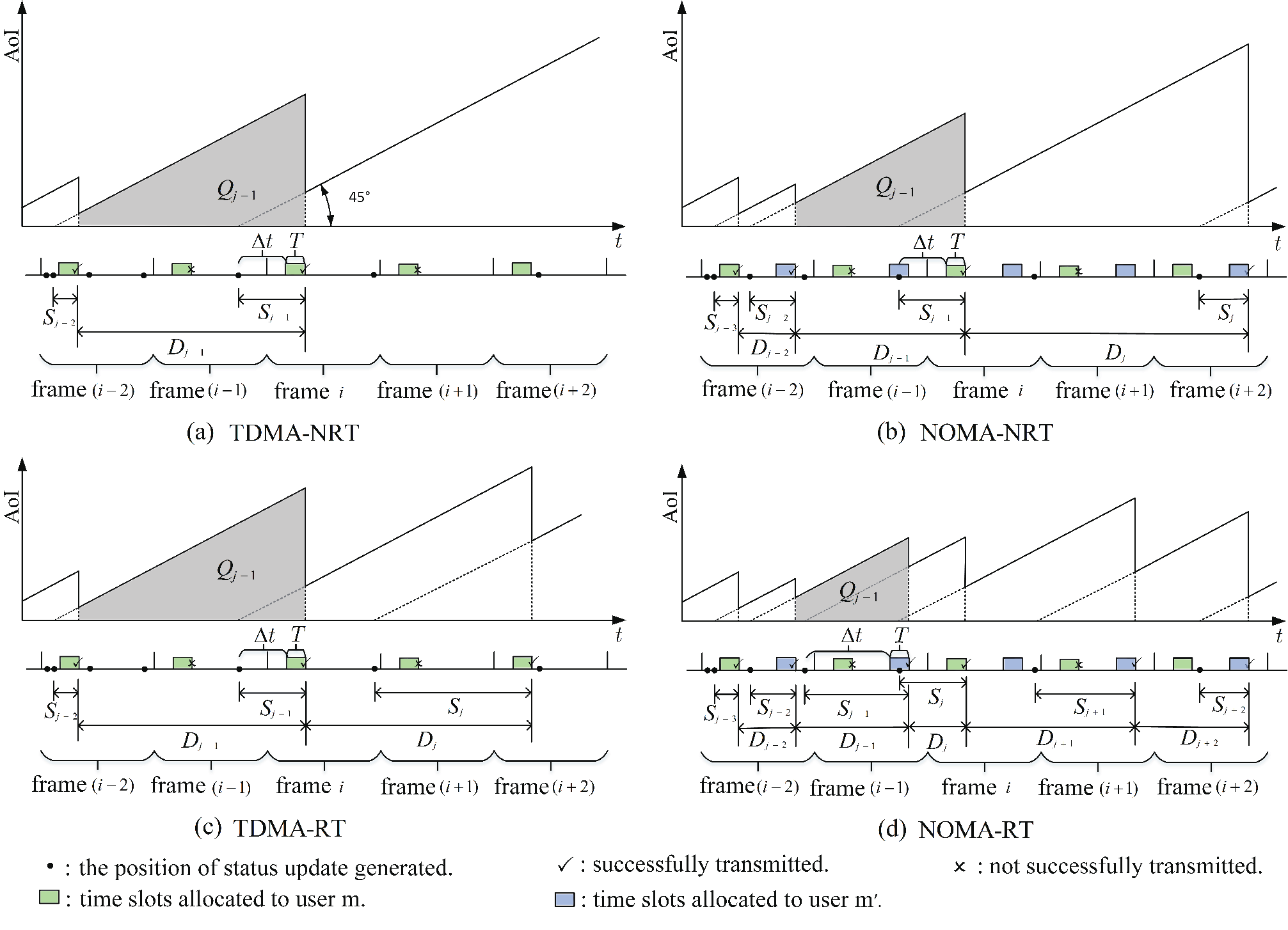}\\
  \caption{Illustration of the status updating process and the corresponding AoI evolution for TDMA-NRT,NOMA-NRT, TDMA-RT and NOMA-RT. It can be seen that by using NOMA and retransmission mechanism, more transmission opportunities can be provided, which can significantly reduce the instantaneous AoI. }\label{update process_exp}
\end{figure*}
\subsection{AoI analysis for TDMA-NRT}
The average AoI achieved by the considered TDMA-NRT scheme can be characterized by the following theorem.
\begin{Theorem}
The average AoI achieved by the considered TDMA-NRT scheme, denoted by $\bar{\Delta}_{m}^{TDMA-NRT}$, can be expressed as:
\begin{align}
 &\bar{\Delta}_{m}^{TDMA-NRT}=\Gamma+\frac{MT(2-(1-e^{-\lambda_mMT})P_{mm})}{2(1-e^{-\lambda_mMT})P_{mm}},
\end{align}
where $\Gamma=\frac{MT}{1-e^{\lambda_mMT}}+\frac{1}{\lambda_m}+T$.
\end{Theorem}
\begin{IEEEproof}
Please refer to Appendix A.
\end{IEEEproof}
\subsection{AoI analysis for NOMA-NRT}
{\color{black}For the TDMA-NRT scheme, the status updating process can be divided into statistically identical and
independent phases with equal duration $MT$. Hence, the derivation of the average AoI for TDMA-NRT can be significantly
simplified by utilizing  the aforementioned property. Different from the TDMA-NRT scheme, where each source $m$ has only
one chance to transmit in a frame, the NOMA-NRT scheme offers an additional transmission chance for source
$m$ by using the $m'$-th slot. As a result, the status updating process in NOMA-NRT can be divided into consecutive
phases with duration $\frac{MT}{2}$. Although it can be easily proved that the status updatings in these phases are
also statistically independent, the probabilities of a successful update in the $m$-th and $m'$-th slot in a frame
are different. As a consequence, it is necessary to take into account where a successful updates happens for the derivation of the average AoI, which is the main challenge caused by NOMA-NRT compared to TDMA-NRT.} The average AoI achieved by the considered NOMA-NRT scheme can be characterized by the following theorem.
\begin{Theorem}
The average AoI achieved by the considered NOMA-NRT scheme, denoted by $\bar{\Delta}_{m}^{NOMA-NRT}$, can be expressed as:
\begin{align}
 \bar{\Delta}_{m}^{NOMA-NRT}=&\bar{\Omega}+\frac{MT(A+B)}{4C},
\end{align}
where $\bar{\Omega}=\frac{MT}{2(1-e^{\frac{\lambda_mMT}{2}})}+\frac{1}{\lambda_m}+T$, $A=p_1p_2(p_0^2+6p_0+1)$, $B=2(1+p_0)(p_1^2(1-p_2)+{p_2}^2(1-p_1))$, $C=(p_1p_2(1+p_0-p_1-p_2)+{p_1}^2+{p_2}^2)(1-p_0)$, $p_0=(1-p_1)(1-p_2)$, $p_1=(1-e^{-\frac{\lambda_{m}MT}{2}})P_{mm}$, and
$
p_2=(1-e^{-\frac{\lambda_{m}MT}{2}})\left((1 -e^{-\frac{\lambda_{m'}MT}{2}})
(\frac{e^{-\frac{\varepsilon}{P^s}}}{1+\frac{P\varepsilon}{P^s}})\!+\!(e^{-\frac{\lambda_{m'}MT}{2}})(e^{-\frac{\varepsilon}{P^s}})\right).$
\end{Theorem}
\begin{IEEEproof}
Please refer to Appendix B.
\end{IEEEproof}
\subsection{AoI analysis for TDMA-RT}
The average AoI achieved by the considered TDMA-RT scheme can be characterized by the following theorem.
\begin{Theorem}
The average AoI achieved by the considered TDMA-RT scheme, denoted by $\bar{\Delta}_{m}^{TDMA-RT}$, can be expressed as:
\begin{align}
\bar{\Delta}_{m}^{TDMA-RT}=\Gamma+\frac{MT(1-P_{mm})e^{-\lambda_mMT}}{1-(1-P_{mm})e^{-\lambda_mMT}}+\frac{\Lambda}{2\Psi},
\end{align}
where $\Lambda=M^2T^2(\frac{(1+P_{mm}-e^{-\lambda_mMT})(3e^{-\lambda_mMT}-1)}{(1-e^{-\lambda_mMT})^2P_{mm}}\!\!+\!\!\frac{2+{P_{mm}}^2}{{P_{mm}}^2})$ , $\Psi=\frac{MT(1+P_{mm}e^{-\lambda_mMT}-e^{-\lambda_mMT})}{(1-e^{-\lambda_mMT})P_{mm}}$.
\end{Theorem}
\begin{IEEEproof}
Please refer to Appendix C.
\end{IEEEproof}

\begin{Remark}
{\color{black}The main difference between the analysis for TDMA-NRT and TDMA-RT schemes is that there's time
correlation of the status updating process in TDMA-RT. Because whether there's packet to transmit in the current slot
depends on not only the arrival of new packet in the $\frac{M}{2}$ interval, but also whether there's packet
which was not successfully transmitted in the last frame. By using tools from the Markov chain theory,
the average AoI achieved by TDMA-RT can be obtained, as highlighted in the following theorem.}
\end{Remark}
\vspace{-3mm}
\subsection{AoI analysis for NOMA-RT}
\begin{figure*}[!t]
\centering
\small
\begin{align} &Y\!\!=\!\!\frac{MT(e^{-\frac{\lambda_mMT}{2}}(1\!+\!P_{m0}(e^{-\lambda_mMT}))(1\!\!-\!\!P_{mm'})\!+\!2P_{m0}e^{-\lambda_mMT}\!)}{2(1\!\!-\!\!P_{m0}(e^{-\lambda_mMT}))(1\!+\!(1-P_{mm'})e^{-\frac{\lambda_mMT}{2}})}, Y'\!\!=\!\!\frac{MT(e^{-\frac{\lambda_mMT}{2}}(1\!+\!P_{m0}(e^{-\lambda_mMT}))(1\!\!-\!\!P_{mm})\!+\!2P_{m0}e^{-\lambda_mMT}\!)}{2(1\!\!-\!\!P_{m0}(e^{-\lambda_mMT}))(1\!+\!(1-P_{mm})e^{-\frac{\lambda_mMT}{2}})}.\label{eq:equation1}\\
&W\!=\!\frac{MT}{2}\!(\!\frac{2(1\!+\!e^{-\frac{\lambda_mMT}{2}})\!-\!P_{mm'}\!-\!e^{-\frac{\lambda_mMT}{2}}P_{mm}}{(1\!-\!P_{m0})(1\!+\!e^{-\frac{\lambda_mMT}{2}})}\!\!+\!\!\frac{e^{-\frac{\lambda_mMT}{2}}}{1\!\!-\!e^{-\frac{\lambda_mMT}{2}}}), W'\!=\!\frac{MT}{2}\!(\!\frac{2(1\!+\!e^{-\frac{\lambda_mMT}{2}})\!-\!P_{mm}\!-\!e^{-\frac{\lambda_mMT}{2}}P_{mm'}}{(1\!-\!P_{m0})(1\!+\!e^{-\frac{\lambda_mMT}{2}})}\!\!+\!\!\frac{e^{-\frac{\lambda_mMT}{2}}}{1\!\!-\!e^{-\frac{\lambda_mMT}{2}}}).\label{eq:equation2}\\
&H\!=\!\frac{M^2T^2}{4}(\frac{\gamma_1(e^{-\frac{3\lambda_mMT}{2}}L+e^{-\lambda_mMT}R+e^{-\frac{\lambda_mMT}{2}}S+T)}{(\gamma_1+\gamma_4)(1-e^{-\lambda_mMT})^2(1-P_{m0})^2}+\frac{\gamma_4(e^{-\frac{3\lambda_mMT}{2}}L'\!+\!e^{-\lambda_mMT}R'\!+\!e^{-\frac{\lambda_mMT}{2}}S'+T')}{(\gamma_1+\gamma_4)(1-e^{-\lambda_mMT})^2(1-P_{m0})^2}).\label{eq:equation3}
\end{align}
\vspace{-2em}
\end{figure*}
{\color{black}The analysis for NOMA-RT scheme is the most challenging among the considered four schemes. The reasons
are mainly as follows:
\begin{itemize}
  \item there's time correlation for the status updating process of a single source. Because, whether there's packet
  to transmit in the current slot is affected by the updating status of the last transmission slot.
  \item there's also correlation between paired sources. The reason is that the transmission success of a
        secondary user depends on the primary user's transmission.
  \item The single-user time correlation and the inter-user correlation are coupled as illustrated by Fig. \ref{delta_PP}.
\end{itemize}

To derive the average AoI achieved by source $m$ in NOMA-RT, it is necessary to first derive the expression for
the transmission success probability $P_{mm'}$ under the stationary state of the status updating process.}

\begin{Lemma}
For the NOMA-RT scheme, conditioning on the steady state of the status updating process, the transmission success probability $P_{mm'}$ when $U_m$ transmits in the $m'$-th time slot, can be expressed as the solution of the following equation:
\begin{align}\label{NOMA-RT-Pmm'}
\hat{a}P_{mm'}^2+\hat{b}P_{mm'}+\hat{c}=0,
\end{align}
where $\Theta=\frac{e^{-\frac{\varepsilon}{P^s}}}{1+\frac{P\varepsilon}{P^s}}$, and
\begin{align}\label{}
\hat{a}=&(a_2\Theta+b_2e^{\frac{-\varepsilon}{P^s}})c_1+c_2d_1,\\
\hat{b}=&(c_1P_{mm}e^{-\lambda_mMT}\!-\!c_2P_{m'm'}e^{-\lambda_{m'}MT})e^{-\frac{\varepsilon}{P^s}}+\\\notag
&\!(c_1(1\!\!-\!e^{-\lambda_{m}MT}\!)\!-\!c_2(1\!\!-\!e^{-\lambda_{m'}MT}\!)\!\!-\!\!(a_2b_1\!\!+\!a_1b_2)e^{-\frac{\varepsilon}{P^s}})\Theta\\\notag
&-a_1a_2\Theta^2\!-\!b_1b_2e^{-\frac{2\varepsilon}{P^s}}\!+\!d_1d_2,\\
\hat{c}=&-d_2((1-e^{-\lambda_{m'}MT})\Theta+P_{m'm'}e^{-\lambda_{m'}MT}e^{-\frac{\varepsilon}{P^s}})\\\notag
&\!\!-\!\!(a_1\Theta\!+\!b_1e^{-\frac{\varepsilon}{P^s}})((1\!\!-\!e^{-\lambda_mMT})\Theta\!+\!P_{mm}e^{-\lambda_mMT}e^{-\frac{\varepsilon}{P^s}}),
\end{align}
and $P_{m'm'}\!=\!P_{mm}\!=\!e^{-\frac{\varepsilon}{P}}$, $a_1\!=\!(e^{-\frac{\lambda_{m'}MT}{2}}\!-\!1)e^{-\frac{\lambda_{m'}MT}{2}}$, $b_1\!=\!(1\!-\!P_{m'm'}e^{-\frac{\lambda_{m'}MT}{2}})e^{-\frac{\lambda_{m'}MT}{2}}$, $a_2\!=\!(e^{-\frac{\lambda_{m}MT}{2}}\!-\!1)e^{-\frac{\lambda_{m}MT}{2}}$, $b_2\!=\!(1\!-\!P_{mm}e^{-\frac{\lambda_{m}MT}{2}})e^{-\frac{\lambda_{m}MT}{2}}$,
$c_1\!=\!a_1\!+\!b_1$, $c_2\!=\!a_2\!+\!b_2$,
$d_1\!=\!(1\!-\!e^{-\lambda_{m'}MT})\!+\!e^{-\lambda_{m'}MT}P_{m'm'}$, $d_2\!=\!(1\!-\!e^{-\lambda_{m}MT})\!+\!e^{-\lambda_{m}MT}P_{mm}$.
\end{Lemma}
\begin{IEEEproof}
Please refer to Appendix D.
\end{IEEEproof}

\begin{Theorem}
The average AoI achieved by the considered NOMA-RT scheme, denoted by $\bar{\Delta}_m^{N\!O\!M\!A\!-\!R\!T}$, can be expressed as:
\begin{align}
 &\bar{\Delta}_m^{N\!O\!M\!A\!-\!R\!T}\!\!=\!\!
 \frac{\gamma_1(\bar{\Omega}\!+\!Y)W\!\!+\!\!\gamma_4(\bar{\Omega}\!+\!Y')W'}{\gamma_1 W\!+\!\gamma_4W'}\!+\!\frac{H(\gamma_1\!+\!\gamma_4)}{2(\gamma_1W\!+\!\gamma_4W')},
\end{align}
where
\begin{align}\label{}
\gamma_1=&\frac{P_{mm}(e^{\frac{\lambda_mMT}{2}}-1)(e^{\frac{\lambda_mMT}{2}}-P_{mm'}+1)}{2(P_{mm}+P_{mm'}+e^{\lambda_mMT}-P_{mm}P_{mm'}-1)} \label{eq:gamma1},\\
\gamma_4=&\frac{P_{mm'}(e^{\frac{\lambda_mMT}{2}}-1)(e^{\frac{\lambda_mMT}{2}}-P_{mm}+1)}{2(P_{mm}+P_{mm'}+e^{\lambda_mMT}-P_{mm}P_{mm'}-1)} \label{eq:gamma2}.
\end{align}
and $Y$, $Y'$, $W$, $W'$ and $H$ are shown in (\ref{eq:equation1}), (\ref{eq:equation2}) and (\ref{eq:equation3}), respectively.
\end{Theorem}
\begin{IEEEproof}
Please refer to Appendix E.
\end{IEEEproof}

\section{Numerical Results}
In this section, numerical results are presented to verify the accuracy of the developed analysis, and also demonstrate AoI performance achieved by the considered TDMA-NRT, TDMA-RT, NOMA-NRT and NOMA-RT schemes.
\begin{figure}[!h]
  \centering
	\setlength{\abovecaptionskip}{0em}   
	\setlength{\belowcaptionskip}{-2em}
  \subfloat[TDMA-NRT and NOMA-NRT]{\includegraphics[width=3in]{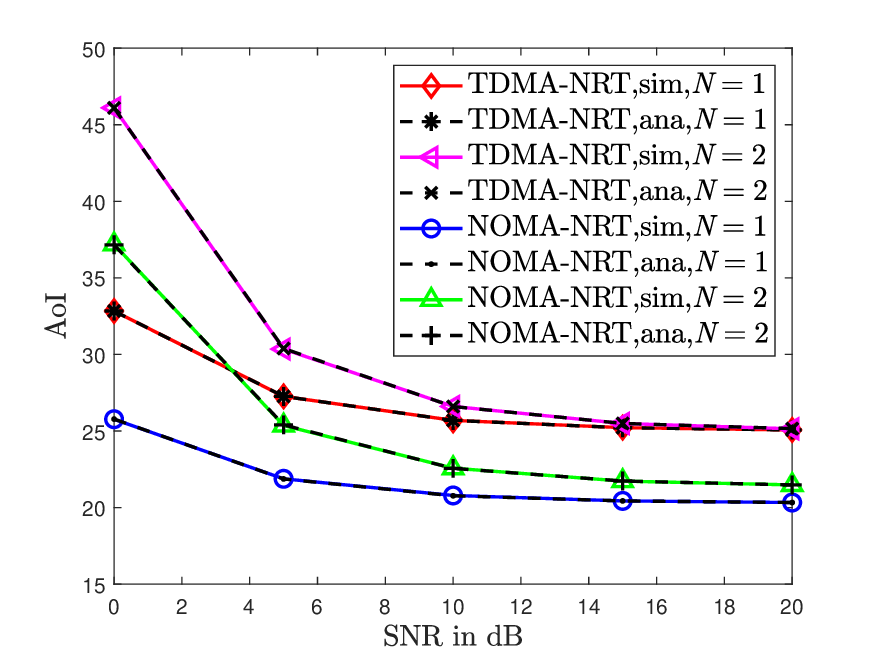}}\\
  \vspace{-1.2em}
  \subfloat[TDMA-RT and NOMA-RT]{\includegraphics[width=3in]{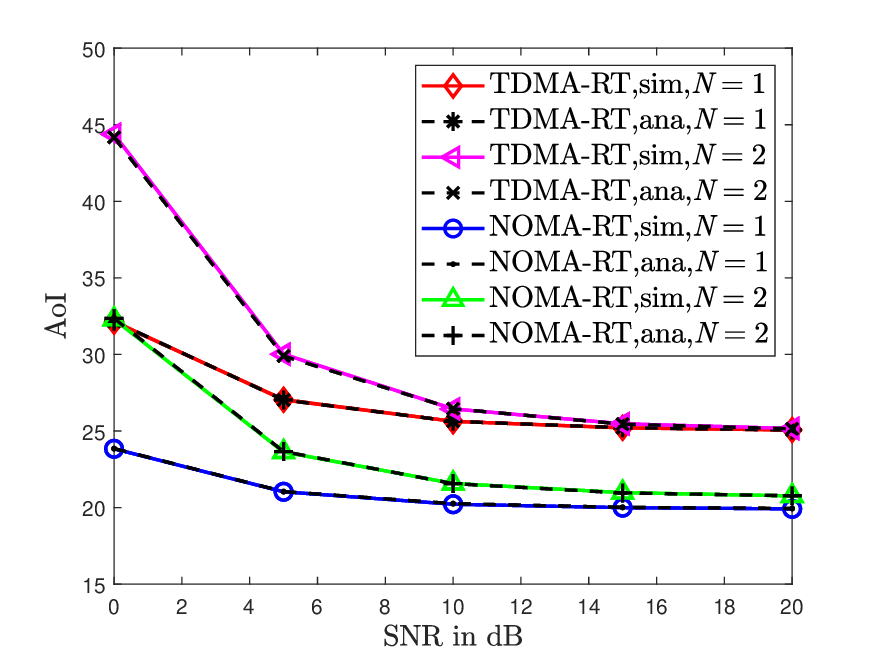}}
  \caption{Average AoI achieved by TDMA-NRT, TDMA-RT, NOMA-NRT and NOMA-RT.
$\lambda_m=\lambda_{m'}=0.1$, $M=8$, $T=3$.}\label{accuracy}
\end{figure}

Fig. \ref{accuracy} shows the average AoI achieved by TDMA-NRT, TDMA-RT, NOMA-NRT and NOMA-RT schemes.
The simulations results are obtained by averaging over $10^5$ consecutive frames. It can be clearly observed from both Fig. \ref{accuracy}(a) and Fig. \ref{accuracy}(b), simulation results perfectly match the analytical results for all the considered schemes, which validates the accuracy of the developed analysis.
Besides, it can be seen from Fig. \ref{accuracy}(a) and Fig. \ref{accuracy}(b) that the average AoIs achieved by NOMA-NRT and NOMA-RT schemes outperform their TDMA counterparts.
\begin{figure}[!t]
  \centering
  	\centering
	\setlength{\abovecaptionskip}{0em}   
	\setlength{\belowcaptionskip}{-2em}
    \subfloat[TDMA-NRT and NOMA-NRT]{\includegraphics[width=3in]{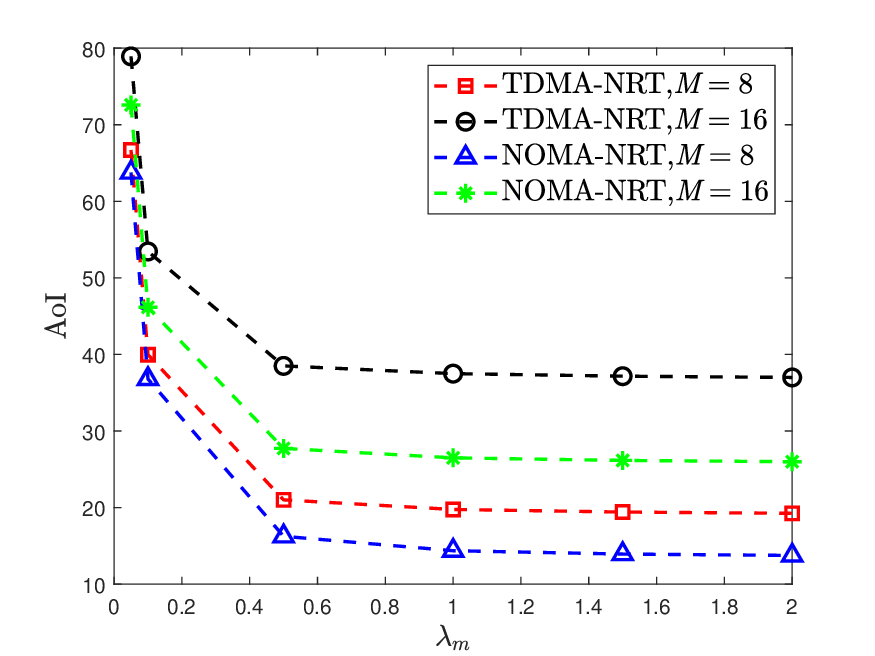}}\\
    \vspace{-1.2em}
  \subfloat[TDMA-RT and NOMA-RT]{ \includegraphics[width=3in]{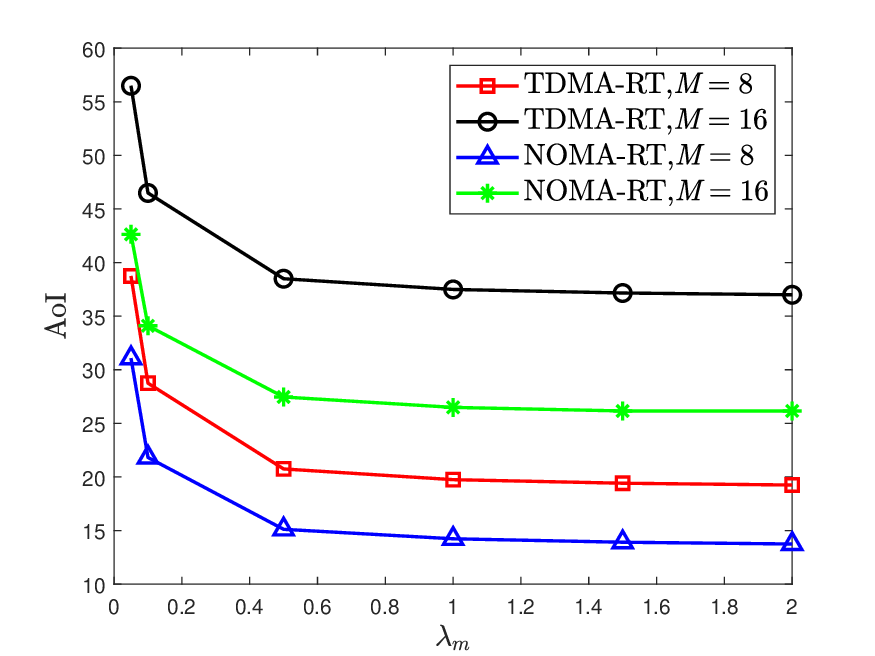}}
  \caption{Impact of packet arrival rates on AoI for TDMA-NRT, TDMA-RT, NOMA-NRT and NOMA-RT. $N=1$ bit, $T=1$, $\lambda_m=\lambda_{m'}$.}\label{impact_lambda}
\end{figure}

Fig. \ref{impact_lambda} demonstrates the impact of the packet arrival
rate on the average AoI achieved by TDMA-NRT, TDMA-RT, NOMA-NRT and NOMA-RT schemes, respectively. As shown in the figure, at low arrival rates, the AoIs achieved by TDMA-NRT, TDMA-RT, NOMA-NRT and NOMA-RT decrease rapidly with the increase of arrival rates. In contrast, at high arrival rates, the AoIs achieved by the four schemes approach a constant, respectively. Another interesting observation is that, for both cases with and without retransmission, the gap between the AoIs achieved by CR-NOMA and TDMA at a high arrival rate is much larger than that at a low arrival rate. This is because at a low arrival rate, the AoI is significantly limited by the arrival rate, while at a high arrival rate, the AoI is limited more by the opportunities to transmit status updates.

\begin{figure}[!t]
  \centering
  	\centering
	\setlength{\abovecaptionskip}{0em}   
	\setlength{\belowcaptionskip}{-2em}
    \subfloat[TDMA-NRT and NOMA-NRT]{\includegraphics[width=3in]{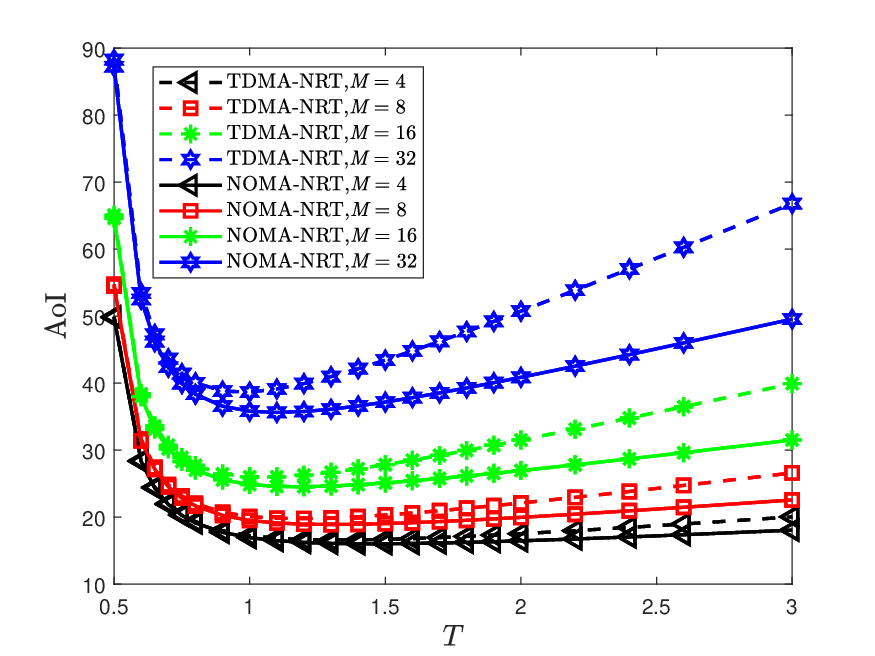}}\\
    \vspace{-1.2em}
    \subfloat[TDMA-RT and NOMA-RT]{\includegraphics[width=3in]{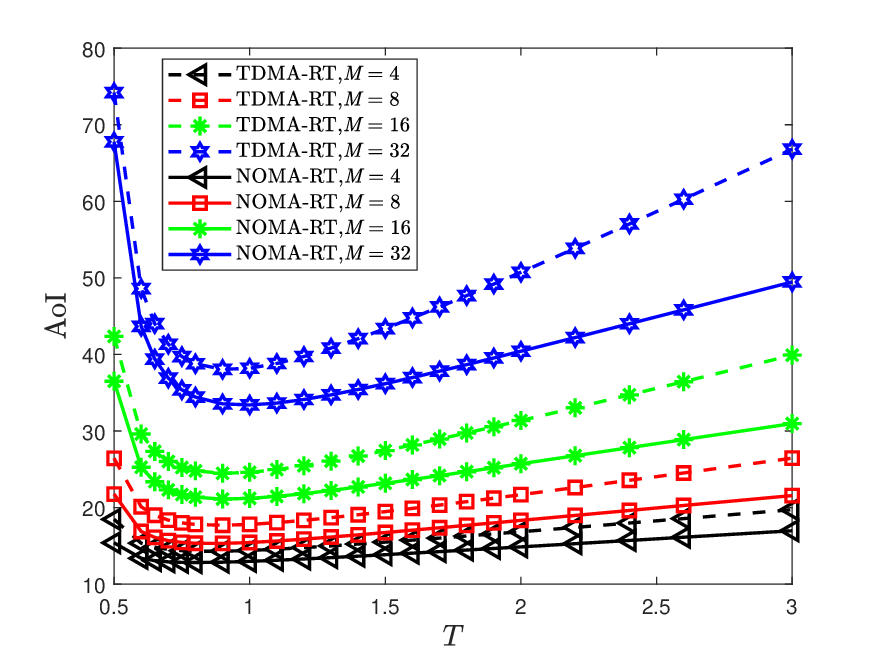}}
  \caption{Impact of the duration of a time slot on AoI for TDMA-NRT, TDMA-RT, NOMA-NRT and NOMA-RT. $N=2$ bits, $\lambda_m=\lambda_{m'}=0.1$.}\label{impact_T}
\end{figure}
\begin{figure}[!h]
  \centering
  	\centering
  	\setlength{\abovecaptionskip}{0em}   
	\setlength{\belowcaptionskip}{-2em}
    \subfloat[TDMA-NRT and NOMA-NRT]{\includegraphics[width=3in]{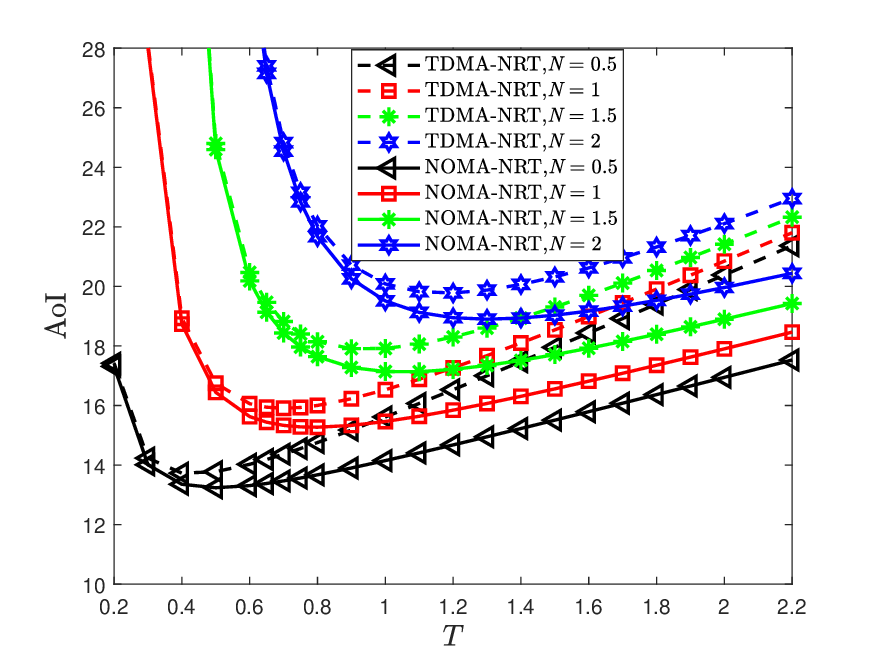}}\\
    \vspace{-1.2em}
  \subfloat[TDMA-RT and NOMA-RT]{ \includegraphics[width=3in]{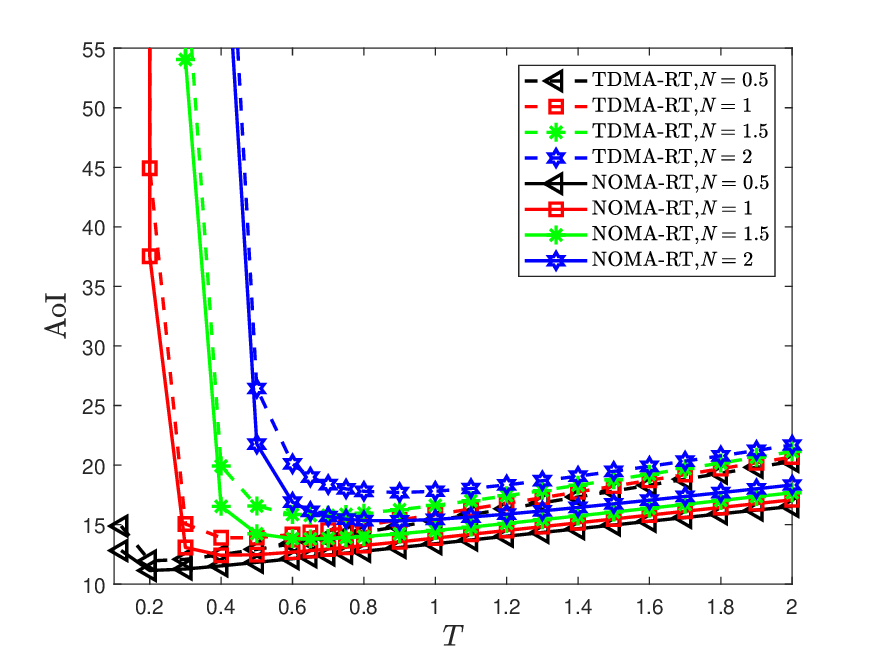}}
  \caption{Impact of the duration of a time slot on AoI for TDMA-NRT, TDMA-RT, NOMA-NRT and NOMA-RT. $M=8$, $\lambda_m=\lambda_{m'}=0.1$.}\label{figadd1}
\end{figure}
\begin{figure}[!h]
  \centering
  	\centering
  	\setlength{\abovecaptionskip}{0em}   
	\setlength{\belowcaptionskip}{-2em}
    \subfloat[TDMA-NRT and NOMA-NRT]{\includegraphics[width=3in]{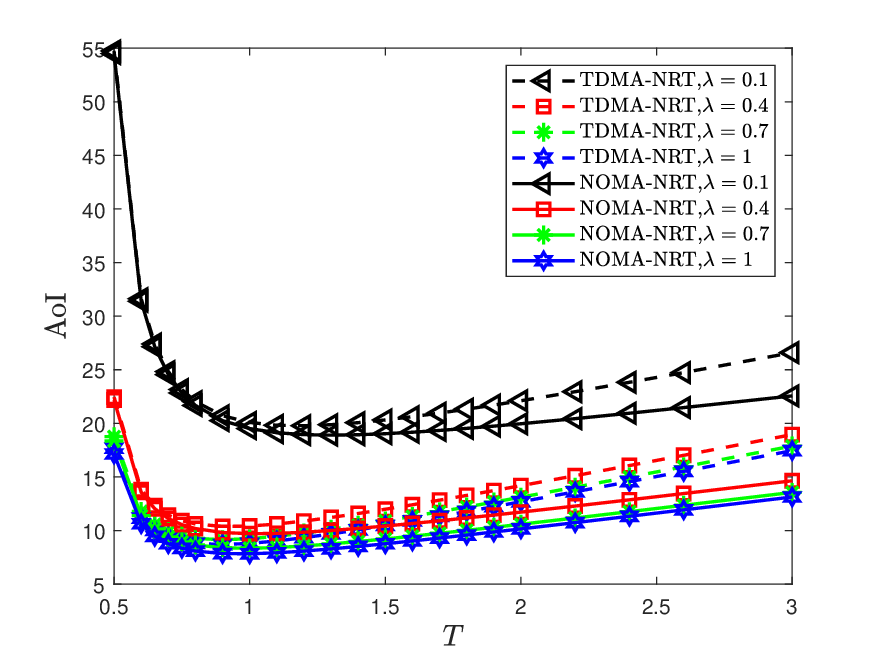}}\\
    \vspace{-1.2em}
  \subfloat[TDMA-RT and NOMA-RT]{ \includegraphics[width=3in]{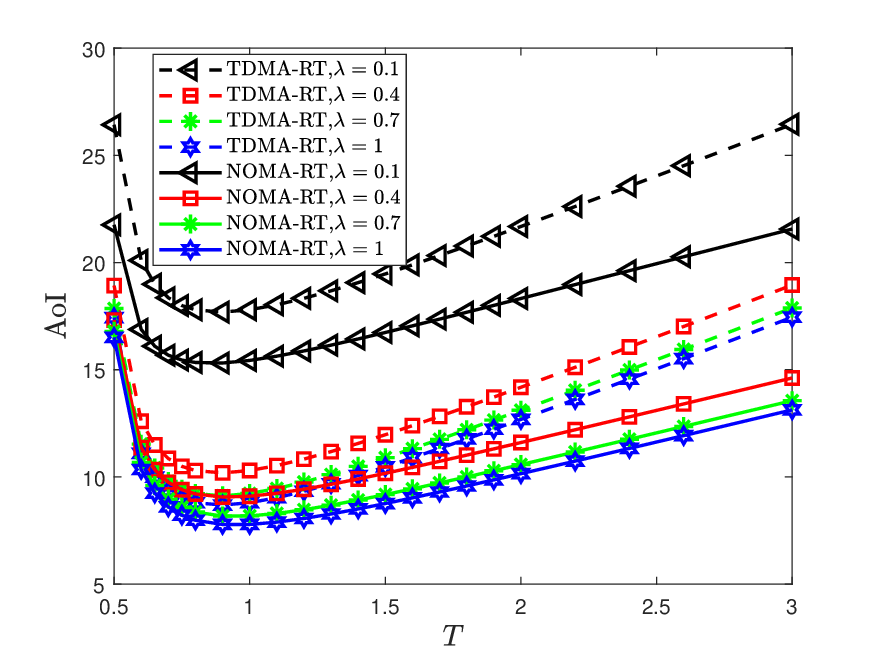}}
  \caption{Impact of the duration of a time slot on AoI for TDMA-NRT, TDMA-RT, NOMA-NRT and NOMA-RT. $N=2$ bits, $M=8$.}\label{figadd2}
\end{figure}
{\color{black}Figs. \ref{impact_T}-\ref{figadd2} show the impact of the duration of a time slot on the AoI achieved by TDMA-NRT, TDMA-RT, NOMA-NRT and NOMA-RT schemes under different values of the number of users,  packet size and data arrival rate, respectively.
As shown in the three figures, for both cases with and without retransmission, the AoIs achieved by both TDMA and CR-NOMA first decrease with $T$ and then increase. This observation can be explained by the following two facts. On the one hand,  as $T$ increases, the frame length will increase, which is unfavorable for reducing AoI. On the other hand, as $T$ increases,
the transmission reliability can be increased, which is beneficial for reducing
AoI. Hence, for a small $T$, the dominant factor for reducing AoI is the transmission reliability, and as a result, increasing $T$ can help to reduce the AoI.
Besides, when $T$ is sufficiently large, the dominant limitation for reducing AoI becomes the length of each frame, and as a result, increasing $T$ yields a larger AoI.
It can also be seen from Figs. \ref{impact_T}-\ref{figadd2} that the optimal value of $T$ is affected by the number of users, the packet sizes and the data arrival rates, due to the fact that these parameters also affect the frame length and transmission reliability. However, the impacts of the aforementioned three factors on the optimal value of $T$ are coupled with each other, which makes it difficult to analyze the optimal value of $T$.}

\begin{figure}[!h]
  \centering
    \setlength{\abovecaptionskip}{0em}   
	\setlength{\belowcaptionskip}{-2em}
  \includegraphics[width=3in]{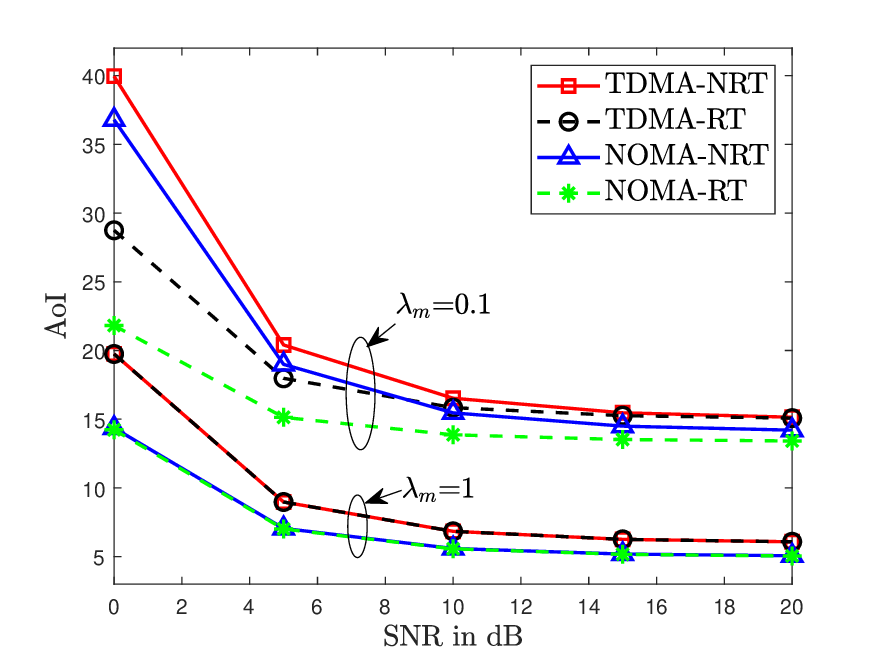}\\
  \caption{Comparisons among TDMA-NRT, TDMA-RT, NOMA-NRT and NOMA-RT in terms of AoI. $M=8$, $N=1$ bit, $T=1$, $\lambda_m=\lambda_{m'}$.}\label{compare}
\end{figure}
\begin{figure}[!h]
  \centering
  	\centering
  	\setlength{\abovecaptionskip}{0em}   
	\setlength{\belowcaptionskip}{-1em}
    \subfloat[$T=1$]{\includegraphics[width=3in]{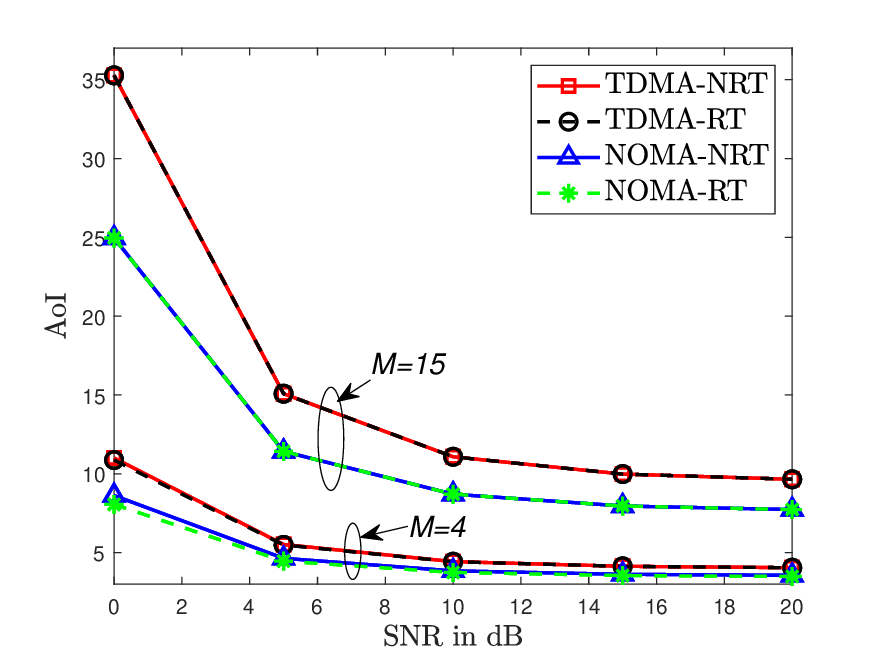}}\\
    \vspace{-1.2em}
  \subfloat[$T=2$]{ \includegraphics[width=3in]{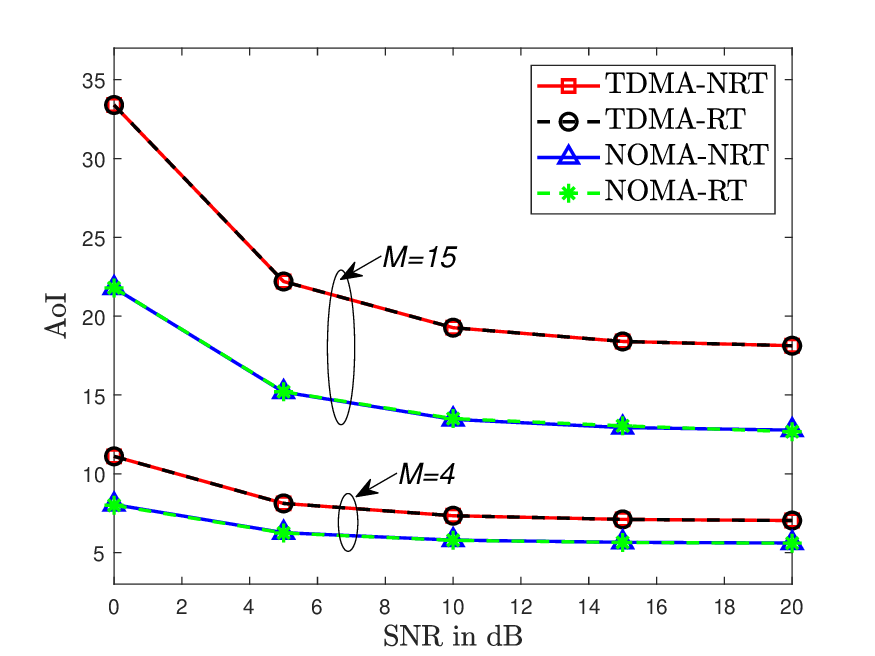}}
  \caption{Comparisons among TDMA-NRT, TDMA-RT, NOMA-NRT and NOMA-RT in terms of AoI. $N=1$ bit, $\lambda_m=\lambda_{m'}=1$.}\label{addSNR}
\end{figure}

{\color{black}Fig. \ref{compare} and Fig. \ref{addSNR}  show the comparison of TDMA-NRT, TDMA-RT, NOMA-NRT and NOMA-RT in terms of average AoI. As shown in Fig. \ref{compare} and \ref{addSNR}, the NOMA-NRT and NOMA-RT schemes outperform the TDMA-NRT and TDMA-RT schemes, respectively. Furthermore, it can be observed that the gap between the AoIs achieved by the NOMA-NRT scheme (or NOMA-RT scheme) and the NOMA-NRT scheme (or NOMA-RT scheme) at a low SNR is larger than that at a high SNR.
In addition, as shown in Fig. \ref{compare}, when $\lambda_m=0.1$, NOMA-RT achieves lower average AoI compared to NOMA-NRT. By contrast, when $\lambda_m=1$,  the curves for NOMA-RT and NOMA-NRT overlaps with each other. Thus, it can be concluded that re-transmission strategy is more necessary for scenarios with low packet arrival rates.
It can also be observed from Fig. \ref{addSNR} that the gap between the AoIs achieved by NOMA-NRT (or NOMA-RT) scheme and TDMA-NRT (or TDMA-RT) scheme increases with the increase of the number of users. The reason can be explained as the following two folds. First, CR-NOMA outperform its TDMA counterpart is mainly because of its shorter waiting time for consecutive transmissions. Second, as the number of users increases, the additional waiting time of TDMA compared to NOMA becomes longer, and hence enlarging the AoI gap.}

\begin{figure}[!h]
  \centering
  	\centering
	\setlength{\abovecaptionskip}{0em}   
	\setlength{\belowcaptionskip}{-1em}
    \subfloat[$\lambda_m=0.5$]{\includegraphics[width=1.8in]{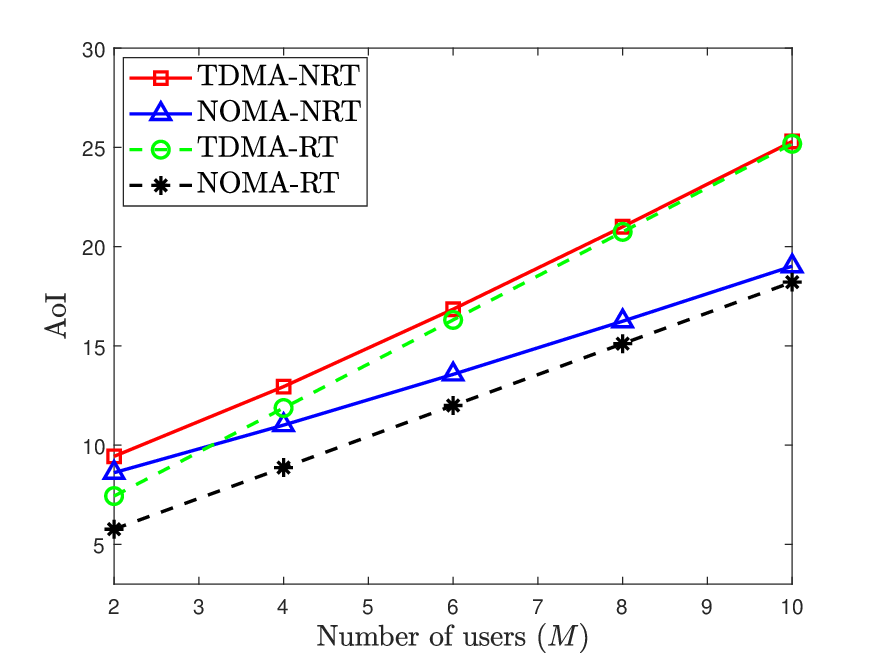}}
  \hspace{-0.65cm}
  \subfloat[$\lambda_m=1$]{ \includegraphics[width=1.8in]{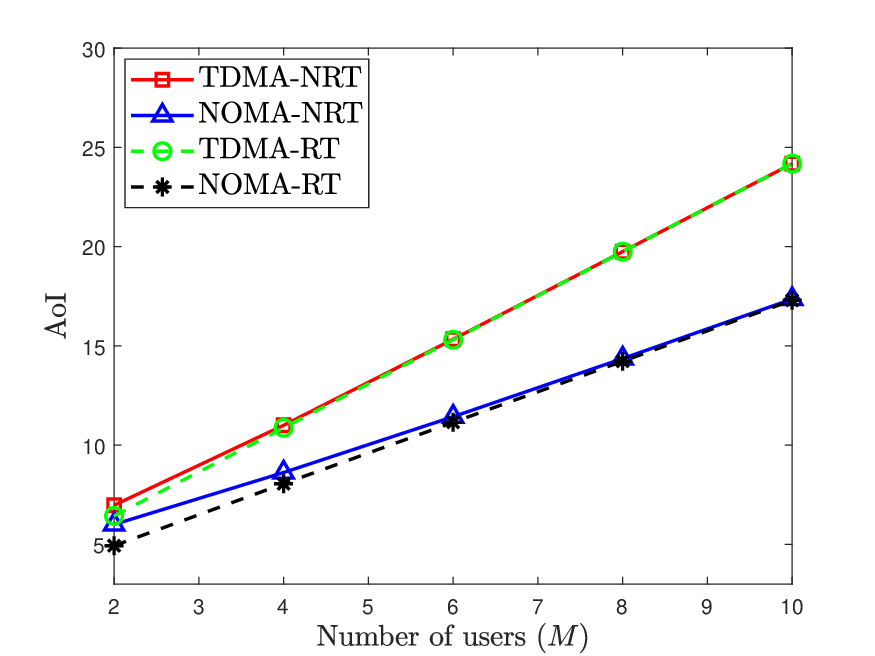}}\\
  \vspace{-1.1em}
  \subfloat[$\lambda_m=1.5$]{ \includegraphics[width=1.8in]{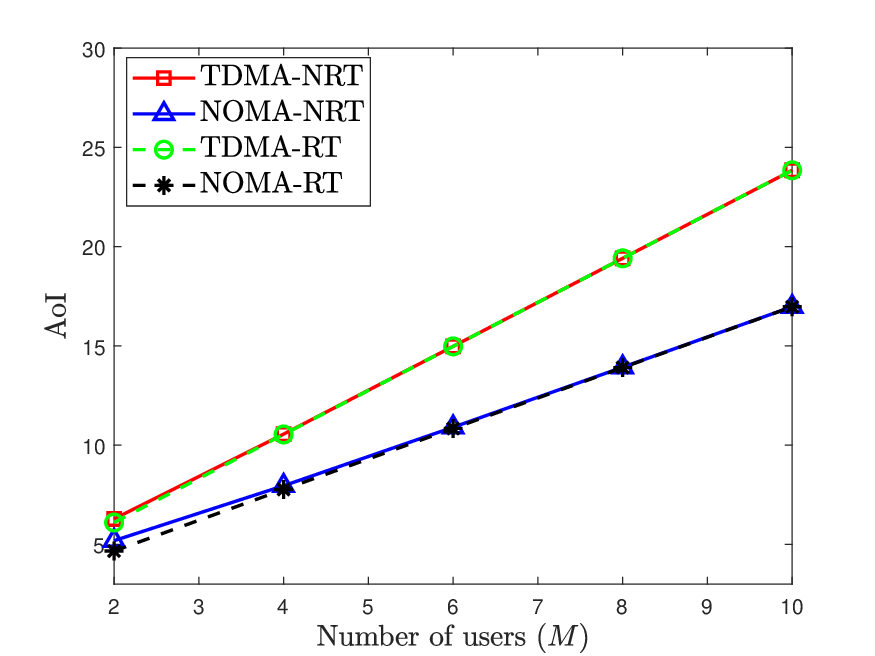}}
  \hspace{-0.65cm}
  \subfloat[$\lambda_m=3$]{ \includegraphics[width=1.8in]{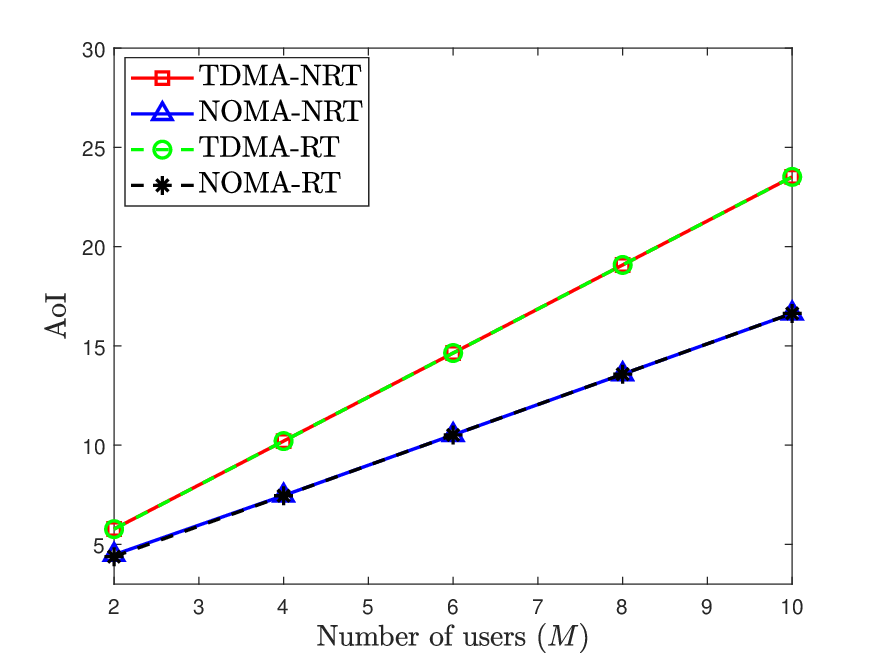}}
  \caption{Impact of the number of users on average AoIs for TDMA-NRT, TDMA-RT, NOMA-NRT and NOMA-RT schemes under different status updating packet arrival rates. $N=1$ bit, $T=1$, $\lambda_m=\lambda_{m'}$.}\label{addM}
\end{figure}
{\color{black}Fig. \ref{addM} shows the impact of the number of users on average AoIs for TDMA-NRT, TDMA-RT, NOMA-NRT and NOMA-RT schemes, under different status updating packet arrival rates. As shown in the figure, the AoIs achieved by TDMA-NRT, TDMA-RT, NOMA-NRT, and NOMA-RT schemes increase with the number of users for a given arrival rate. In addition, the gap between the AoIs achieved by NOMA-NRT (or NOMA-RT) and TDMA-NRT (or TDMA-RT) increases with the number of users. Another interesting observation is that the gap between the AoIs achieved by a retransmission scheme and its corresponding non-retransmission scheme vanishes as the number of users increases. Moreover, as the arrival rate increases, NOMA-NRT scheme achieves almost the same AoI compared to the corresponding NOMA-RT scheme.}
\begin{figure}[!h]
  \centering
    \setlength{\abovecaptionskip}{0em}   
	\setlength{\belowcaptionskip}{-1em}
  \includegraphics[width=3in]{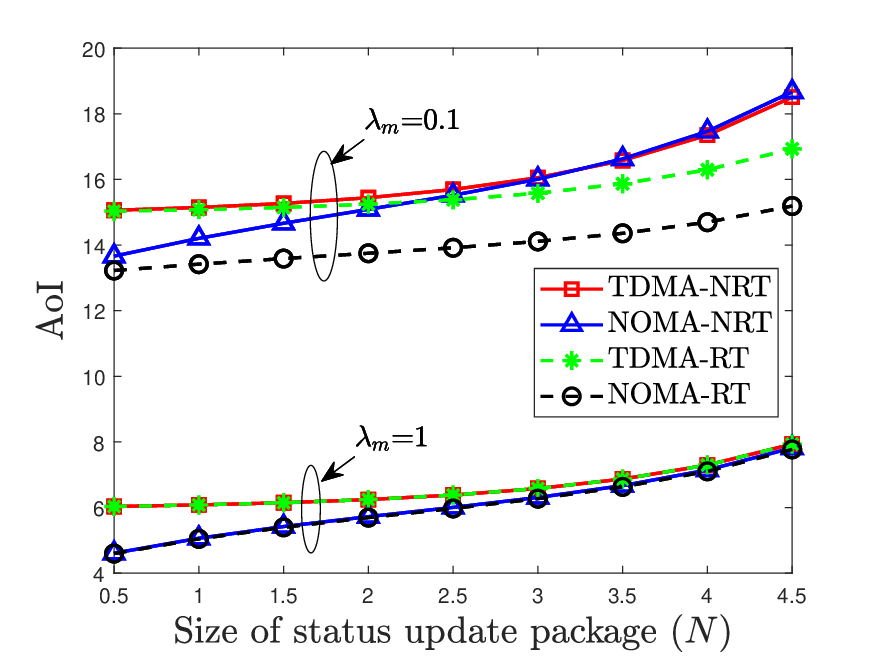}\\
  \caption{Impact of packet size $N$ on average AoI for TDMA-NRT, TDMA-RT, NOMA-NRT and NOMA-RT. $M=8$, $T=1$, SNR$=20$dB.}\label{addNM9}
\end{figure}

{\color{black}Fig. \ref{addNM9} shows the impact of packet size on average AoIs achieved by TDMA-NRT, TDMA-RT, NOMA-NRT and NOMA-RT schemes. As shown in the figure, the AoIs achieved by TDMA-NRT, TDMA-RT, NOMA-NRT, and NOMA-RT increase with the packet size, since larger packet size results in lower transmission reliability.
Interestingly, the comparisons of the curves under different parameter settings behave different. On the one hand, for a low packet arrival rate, both NOMA-RT and NOMA-NRT schemes significantly outperform their TDMA counterparts when the packet size is low. However, NOMA-NRT scheme might not outperform or even be worse than TDMA-NRT scheme when the packet size is relatively large. The reason can be explained as follows. Consider an updating  packet arrives at user $m$ just before user $m'$'s slot, then user $m$ will transmit the packet as a secondary user in the $m'$-th slot. However,  for a large packet size, it is highly possible that the transmission of a secondary user fails. As a consequence, the transmission opportunity is wasted and the packet is dropped, followed by a long waiting time for a new packet's arrival when the arrival rate is low.
On the other hand, for higher packet arrival rates, CR-NOMA schemes outperform their TDMA counterparts. However, as the packet size increases, the performance gain becomes less significant.  This can also be explained by the fact that the transmission reliability of  a secondary user in CR-NOMA decreases as the packet size increases.}

\begin{figure}[!h]
  \centering
    \setlength{\abovecaptionskip}{0em}   
	\setlength{\belowcaptionskip}{-1em}
  \includegraphics[width=3in]{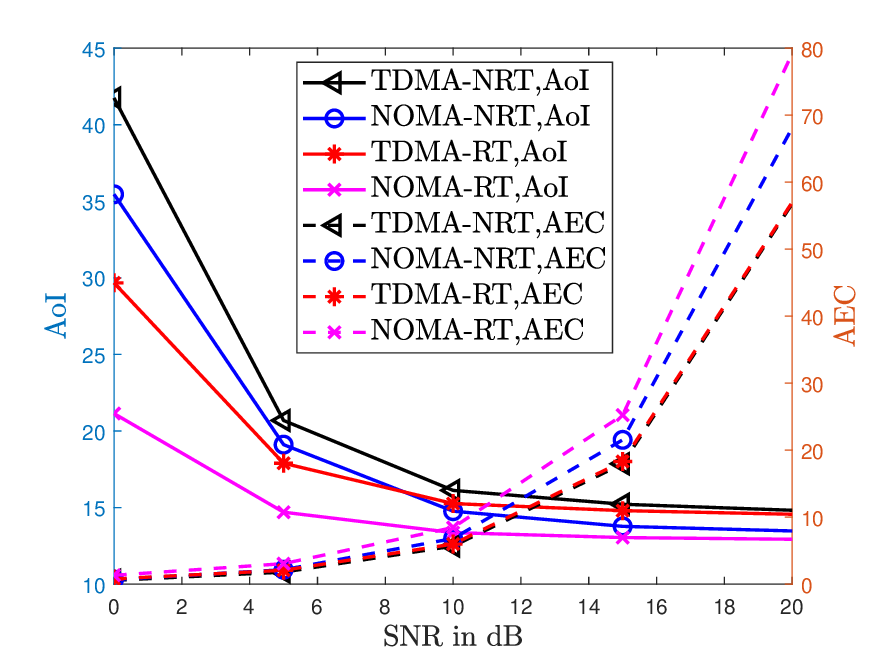}\\
  \caption{Relationship between AoI and AEC. $M=8$, $T=1$, $N=1$ bit, $\lambda_m=\lambda_{m'}=0.1$.}\label{fig}
\end{figure}
{\color{black}Fig. \ref{fig} shows the relationship between average AoI and average energy consumption (AEC) for the TDMA-NRT, TDMA-RT, NOMA-NRT and NOMA-RT schemes. Note that AEC is obtained by the equation:
$\text{AEC}=N_s*P*T/N_a$, where $N_s$ denotes the number of time slots where one source transmits signals, and $N_a$ denotes the total number of frames.
It can be clearly observed from Fig. \ref{fig} that the AoIs decrease and the AECs increase with SNR, and the NOMA-NRT (NOMA-RT) scheme can achieve a smaller AoI than the TDMA-NRT (TDMA-RT) scheme but at the cost of a higher AEC. In addition, NOMA-RT scheme can achieve lower AoI compared to NOMA-NRT scheme, also at the cost of higher AEC. It is important to take the energy budget into consideration when minimizing AoI as an important future research direction.}

\section{Conclusions}
The application of CR-NOMA to reduce information timeliness of status updating systems has been investigated in this paper, where the randomness of the data generation process has been considered. The LCFS queuing strategy has also been adopted. Closed-form expressions for the average AoIs achieved by NOMA-NRT and NOMA-RT schemes have been obtained. Simulation results have been provided to verify the developed analysis and also demonstrate the superior performance of applying CR-NOMA to reduce AoI.

{\color{black}Note that, fixed power allocation has been considered in this paper, considering power budget and designing practical power allocation schemes will be a very important research direction in future. Besides, in this paper, at most two users can transmit in a single time slot. For future work, it is important to study the schemes which can accommodate more users for ensuring the freshness of data in a single slot.}
{\color{black}Moreover, the considered TDMA  has limitations to be applied in dense scenarios for reducing AoI, due to overhead, synchronization issues and  resource allocation complexity.
For scenarios with dense sources, it is important to apply random access methods, such as grant free schemes,  to ensure information freshness, which is left as an important future research direction.}
{\color{black}Last but not least, it can be envisioned that rate splitting multiple access (RSMA) \cite{mao2022rate} has potential to further reduce AoI, by splitting the secondary user's signal into two independent sub-signals, which is left as an important future exploration direction.}

\appendices
\section{Proof for Theorem $1$}
It can be easily found that $D_j$ and $S_{j-1}$ are independent of each other, thus, we have
\begin{align}\label{eq8}
\bar{\Delta}_m^{TDMA-NRT}=\frac{\mathbb{E}\{Q_j\}}{\mathbb{E}\{D_j\}}=\mathbb{E}\{S_{j-1}\}+\frac{\mathbb{E}\{D_j^2\}/2}{\mathbb{E}\{D_j\}}.
\end{align}

\subsubsection{Evaluation of $\mathbb{E}\{S_{j-1}\}$}
as shown in  Fig.\ref{update process_exp} (a), the system time can be divided into two parts as $S_{j-1}=\Delta t+T$,
where $\Delta t$ is the waiting time of the transmitted update since its generation, and $T$ is the transmission time.

It is noteworthy that the evaluation of expectation of the system time is affected by the following two events occur, say $\mathcal{A}_1$ and $\mathcal{A}_2$, where $\mathcal{A}_1$ denotes the event that there is at least one status update generated within the interval with duration $MT$ before the start of the transmission time slot, and $\mathcal{A}_2$ denotes the event that the transmission of the update is successful during the transmission time slot. Thus, the expectation for $S_{j-1}$ can be expressed as
$
 \mathbb{E}\{S_{j-1}\}=\mathbb{E}\{\Delta t| \mathcal{A}\}+T,
$
where $\mathcal{A}=\mathcal{A}_1 \cap \mathcal{A}_2$.

To obtain $\mathbb{E}\{\Delta t| \mathcal{A}\}$, the distribution of $\Delta t$ is required.
The conditional cumulative distribution function (CDF) of $\Delta t$ given $\mathcal{A}$ can be calculated as follows:
 \begin{align}\label{TDMA-NRT CDF}
  F_{\Delta t}(t|\mathcal{A}) &=\frac{ P( \Delta t \le t,\mathcal{A}_1,\mathcal{A}_2) }{P(\mathcal{A}_1,\mathcal{A}_2)}\\\notag
  &\overset{(a)}{=}\frac{ P( \Delta t \le t, \mathcal{A}_2)}{P(\mathcal{A}_1,\mathcal{A}_2)}\\\notag
    &\overset{(b)}{=}\frac{ P( \Delta t \le t)P(\mathcal{A}_2)}{P(\mathcal{A}_1)P(\mathcal{A}_2)}\\\notag
   &\overset{(c)}{=} \frac{1 - e ^{- \lambda_mt}}{1-e^{-\lambda_mMT}}, \quad 0 \le t < MT,
\end{align}
where step (a) follows from the fact that $\Delta t \leq t$ ($0 \le t < MT$) implies the occurrence of $\mathcal{A}_1$,
and step (b) follows from the fact that $\mathcal{A}_2$ is independent of the event that $\Delta t \leq t$ and $\mathcal{A}_1$, respectively, and step (c) is obtained by noting that the status update generation follows a Poisson process.

Then, by taking derivative, the probability density function (PDF) of $\Delta t$ can be obtained as follows:
\begin{align}\label{pdf_deltat}
f_{\Delta t}(t|\mathcal{A})=\frac{\lambda_me^{-\lambda_mt}}{1-e^{-\lambda_mMT}}, \quad 0 \le t < MT.
\end{align}

By applying (\ref{pdf_deltat}), $E(\Delta t|\mathcal{A})$ can be easily obtained as follows:
\begin{align}\label{waiting_time_TDMANRT}
\mathbb{E}\{\Delta t|\mathcal{A}\}\!=\!\int_{0}^{MT} \!t*f_{\Delta t}(t|\mathcal{A})\, dt \!=\!\frac{MT}{1-e^{\lambda_mMT}}\!+\!\frac{1}{\lambda_m}.
 \end{align}

Thus,  $\mathbb{E}\{S_{j-1}\}$ can be expressed as:
 \begin{align}\label{system_time_TDMANRT}
\mathbb{E}\{S_{j-1}\}=\Gamma.
 \end{align}

\subsubsection{Evaluation of $\mathbb{E}\{D_j\}$ and $\mathbb{E}\{D_j^2\}$}
by noting that the status update generating process and transmission for different frames are independent from each other, the evaluation of  $\mathbb{E}\{D_j\}$ and $\mathbb{E}\{D_j^2\}$ can be straightforwardly obtained, as shown in the following.

Note that, the values of $D_j$ can be expressed as $D_j=kMT$, where $k$ denotes any positive integer.
The probability for that $D_j=kMT$ can be expressed as:
\begin{align}
P(D_j\!=\!kMT)\!=\left(1-P(\mathcal{A})\right)^{k-1}P(\mathcal{A}),
\end{align}
where $P(\mathcal{A})=(1\!-\!e^{-\lambda_mMT})P_{mm}$.

Thus, the expression for $\mathbb{E}\{D_j\}$ can be obtained as follows:
\begin{align}\label{eq12}
\mathbb{E}\{D_j\}\!\!=\!\!\!\sum_{k=1}^{\infty}\!(kMT\!)P(\!D_j\!\!=\!\!kMT\!)\!=\!\frac{MT}{(1\!-\!e^{-\lambda_mMT})P_{mm}}.
\end{align}
Similarly, the expression for $\mathbb{E}\{D_j^2\}$ can be obtained as follows:
\begin{align}\label{}
\mathbb{E}\{D_j^2\}=&\sum_{k=1}^{\infty}(kMT)^2P(D_j=kMT)   \\ \notag
=& \frac{M^2T^2(2-(1-e^{-\lambda_mMT})P_{mm})}{(1-e^{-\lambda_mMT})^2P_{mm}^2},
\end{align}
and the proof is complete.

\section{Proof for Theorem $2$}
To obtain the average AoI achieved by NOMA-NRT, the first task is to evaluate $\mathbb{E}\{D_jS_{j-1}\}$.
Note that, the transmission of the $(j-1)$-th successfully delivered update might be finished at the end of either the $m$-th or the $m'$-th time slot of a frame, yielding different distributions of $D_j$. Thus, $\mathbb{E}\{D_jS_{j-1}\}$ can be evaluated as follows:
\begin{align}\label{E(DS)}
&\quad\mathbb{E}\{D_jS_{j-1}\}\\\notag
&=P(G_m)\mathbb{E}\!\{D_jS_{j\!-\!1}|G_m\!\}\!+\!P(G_{m'})\mathbb{E}\!\{D_jS_{j\!-\!1}|G_{m'}\!\}\\\notag
&\overset{(a)}{=}P(G_m)\mathbb{E}\{D_j|G_m\}\mathbb{E}\{S_{j-1}|G_m\}\\\notag
&\quad+P(G_{m'})\mathbb{E}\{D_j|G_{m'}\}\mathbb{E}\{S_{j-1}|G_{m'}\},
\end{align}
where $G_m$ and $G_{m'}$ denote the events that the transmission of the $(j-1)$-th successfully delivered update is finished at the end of the $m$-th and $m'$-th time slot, respectively, and step (a) follows from the fact that, given $G_m$ (or $G_{m'}$), $S_{j-1}$ and $D_j$ are independent of each other.

In the following, it will be shown that the calculation of (\ref{E(DS)}) can be significantly simplified. To this end, we first evaluate $\mathbb{E}\{S_{j-1}|G_m\}$ and $\mathbb{E}\{S_{j-1}|G_{m'}\}$.
As shown in  Fig.\ref{update process_exp} (b), $S_{j-1}$ can be divided into two parts as:
$S_{j-1}=\Delta t+T$, where $\Delta t$ is the waiting time of the transmitted update since its generation, and $T$ is the transmission time.

Note that the evaluation of $S_{j-1}$ should be taken under the condition that the following two events occur, say $\mathcal{B}_1$ and $\mathcal{B}_2$, where $\mathcal{B}_1$ denotes the event that there is  at least one status update generated within the interval with duration $\frac{MT}{2}$ before the start of the
transmission time slot, and $\mathcal{B}_2$ denotes the event that the status update is finally successfully transmitted within the transmitting time slot. It is noteworthy that $\mathcal{B}_2$ can be divided into
two disjoint events, i.e., $\mathcal{B}_2=\mathcal{B}_{2m} \cup \mathcal{B}_{2m'}$, where $\mathcal{B}_{2m}$
and $\mathcal{B}_{2m'}$ denote the transmission is completed within the $m$-th time slot and $m'$-th time slot of a frame, respectively.  Then,  $\mathbb{E}\{S_{j-1}|G_m\}$ and $\mathbb{E}\{S_{j-1}|G_{m'}\}$ can be expressed as follows:
\begin{align}
&\mathbb{E}\{S_{j-1}|G_m\}=\mathbb{E}\{\Delta t|\mathcal{B}_1,\mathcal{B}_{2m}\}+T,\\
&\mathbb{E}\{S_{j-1}|G_{m'}\}=\mathbb{E}\{\Delta t|\mathcal{B}_1,\mathcal{B}_{2m'}\}+T.
\end{align}

In the following, it will be shown how $\mathbb{E}\{S_{j-1}|G_m\}$ can be evaluated.  First, it is necessary to characterize the distribution of $\Delta t$ given $G_m$, which is given by:
\begin{align}\label{distribution}
  F_{\Delta t}(t|G_m)&=\frac{ P( \Delta t \le t,\mathcal{B}_1,\mathcal{B}_{2m}) }{P(\mathcal{B}_1,\mathcal{B}_{2m})}.
\end{align}
$F_{\Delta t}(t|G_m)$ can be calculated as follows:

\begin{align}\label{NOMA CDF}
  F_{\Delta t}(t|G_m)&\overset{(a)}{=}\frac{ P( \Delta t \le t,\mathcal{B}_{2m}) }{P(\mathcal{B}_1,\mathcal{B}_{2m})} \\\notag
    &\overset{(b)}{=}\frac{P( \Delta t \le t)}{P(\mathcal{B}_1)}  \\\notag
     &\overset{(c)}{=} \frac{1 - e ^{- \lambda_mt}}{1-e^{-\lambda_m\frac{MT}{2}}},\quad 0 \le t < \frac{MT}{2},
\end{align}
where step (a) follows from the fact that $\Delta t \leq t$ ($0 \le t < \frac{MT}{2}$) implies the occurrence of $\mathcal{B}_1$, step (b) follows from the fact that $\mathcal{B}_{2m}$ is independent of the event that $\Delta t \leq t$ and $\mathcal{B}_1$, respectively, and  Step (c) is obtained by noting that the status update generation follows a Poisson process.

Then, $E(S_{j-1}|G_m)$ can be easily obtained as follows:
\begin{align}\label{NOMA-NRT E(Sj-1|Gm)}
E(S_{j-1}|G_m)=\int_{0}^{\frac{MT}{2}} \!t*F_{\Delta t}'(t|G_m)\, dt +T =\bar{\Omega}.
 \end{align}
Similarly, $E(S_{j-1}|G_{m'})$ can be expressed as:
\begin{align}\label{E(S/Gm')}
\mathbb{E}\{S_{j-1}|G_{m'}\}=\bar{\Omega}.
 \end{align}
Interestingly, it can be easily found that
 \begin{align}
\mathbb{E}\{S_{j-1}|G_m\}=\mathbb{E}\{S_{j-1}|G_{m'}\},
 \end{align}
which straightforwardly results in
\begin{align}
\mathbb{E}\{S_{j-1}\}=\mathbb{E}\{S_{j-1}|G_m\}=\mathbb{E}\{S_{j-1}|G_{m'}\}.
\end{align}
Thus, $\mathbb{E}\{D_jS_{j-1}\}$ can be further expressed as:
\begin{align}\label{E}
&\mathbb{E}\{D_jS_{j-1}\}\\\notag
=&\mathbb{E}\{S_{j-1}\}(P(G_m)\mathbb{E}\{D_j|G_m\}+P(G_{m'})\mathbb{E}\{D_j|G_{m'}\})\\\notag
=&\mathbb{E}\{S_{j-1}\}\mathbb{E}\{D_j\}.
\end{align}

Hence, $ \bar{\Delta}_{m}^{N\!O\!M\!A\!-\!N\!R\!T}$ can be simplified as follows:
\begin{align}\label{}
 \bar{\Delta}_{m}^{N\!O\!M\!A\!-\!N\!R\!T}=\frac{\mathbb{E}\{Q_j\}}{\mathbb{E}\{D_j\}}=\mathbb{E}\{S_{j-1}\}+\frac{\mathbb{E}\{D_j^2\}/2}{\mathbb{E}\{D_j\}}.
\end{align}

Therefore, the remainder of the proof is to evaluate $\mathbb{E}\{D_j\}$ and $\mathbb{E}\{D_j^2\}$.

To obtain $\mathbb{E}\{D_j\}$ and $\mathbb{E}\{D_j^2\}$, it is necessary to first evaluate the transmission success
probability of $U_m$. The transmission success probability of $U_m$ if the $m$-th time slot is used is
given by $P_{mm}$ as shown in (\ref{P_{mm}}). In contrast, when $U_m$ transmits signal in the $m'$-th time slot, its transmission is likely to be interfered by $U_{m'}$, depending on whether $U_{m'}$ transmits data in the $m'$-th time slot. Hence, the corresponding transmission success probability $P_{mm'}$ can be evaluated as follows:
\begin{align}
P_{mm'}\!=&P(\delta_{i,m'}^{m'}=1)P( R_{i,m'}^{m}\!>\!\frac{N}{T}|\delta_{i,m'}^{m'}=1)\\\notag
&+P(\delta_{i,m'}^{m'}=0)P(R_{i,m'}^m>\frac{N}{T}|\delta_{i,m'}^{m'}=0)\\\notag
=&(1\!-\!e^{-\frac{\lambda_{m'}MT}{2}})(\frac{e^{-\frac{\varepsilon}{P^s}}}{1+\frac{P\varepsilon}{P^s}})\!+\!(e^{-\frac{\lambda_{m'}MT}{2}})(e^{-\frac{\varepsilon}{P^s}}),
\end{align}
where $P(\delta_{i,m'}^{m'}=1)=(1-e^{-\frac{\lambda_{m'}MT}{2}})$.

As aforementioned, the distribution of $D_j$ is dependent on where the last successful update ends, or equivalently,
$G_m$ or $G_{m'}$ happens. Thus, $\mathbb{E}\{D_{j}\}$ can be expressed as:
\begin{align}
\mathbb{E}\{D_{j}\}\!=\!
P\left(G_m\right)\mathbb{E}\{D_{j}|G_m\}+P\left(G_{m'}\right)\mathbb{E}\{D_{j}|G_{m'}\}.
\end{align}

For notational simplicity, denote $p_1$ by the probability of the event that there is status update to be transmitted before the $m$-th time slot of a given frame and it is successfully delivered by using the $m$-th time slot. Similarly, denote $p_2$ as the probability of the event that there is status update to be transmitted before the $m'$-th time slot of a given frame and it is successfully delivered by using the $m'$-th time slot.
It is straightforward to show that $p_1$ and $p_2$ can be expressed as follows:
\begin{align}\label{p1p2}
p_1=(1-e^{-\frac{\lambda_{m}MT}{2}})P_{mm}, p_2=(1-e^{-\frac{\lambda_{m}MT}{2}})P_{mm'}.
\end{align}
Then, $P(G_m)$ and $P(G_{m'})$ can be expressed as:
\begin{align}
 P(G_m)=\frac{p_1}{(p_1+p_2)},P(G_{m'})=\frac{p_2}{(p_1+p_2)}.
\end{align}

To evaluate $\mathbb{E}\{D_j|G_m\}$, it is necessary to characterize the conditional distribution of $D_j$ given $G_m$. Note that the value of $D_j$ can be expressed as $D_j=\frac{kMT}{2}$, where $k$ is a random positive integer. It can be obtained that:
\begin{align}
 P(D_j=\frac{kMT}{2})=
 \begin{cases}
     \!(1-p_2)^{\frac{k}{2}}(1-p_1)^{\frac{k}{2}-1}p_1, k \text{ is even}\\
     \!(1-p_2)^{\frac{k-1}{2}}(1-p_1)^{\frac{k-1}{2}}p_2, k \text{ is odd}
    \end{cases}
\end{align}
Thus, $\mathbb{E}\{D_{j}|G_m\}$ can be obtained as follows:
\begin{align}\label{D_{j}|G_m}
&\mathbb{E}\{D_{j}|G_m\}\\\notag
=&\sum_{k=1}^{\infty}(2k\!-\!1)\frac{MT}{2}p_0^{k\!-\!1}p_2+\sum_{k=1}^{\infty}(2k)\frac{MT}{2}p_0^{k-1}(1\!-\!p_2)p_1\\\notag
=&\frac{MT(2p_1\!+\!p_2\!+\!p_2p_0\!-\!2p_1p_2)}{2(1\!-\!p_0)^2}.
\end{align}
Similarly, $\mathbb{E}\{D_{j}|G_{m'}\}$ can be obtained as follows:
\begin{align}\label{D_{j}|G_m'}
\mathbb{E}\{D_{j}|G_{m'}\}=&\frac{MT(2p_2+p_1+p_1p_0-2p_1p_2)}{2(1-p_0)^2}.
\end{align}

Therefore, with some algebraic manipulations, the expression for $\mathbb{E}\{D_{j}\}$ can be obtained as follows:
\begin{align}
\mathbb{E}\{D_{j}\}
=&\frac{MT\left(p_1p_2(1+p_0-p_1-p_2)+{p_1}^2+{p_2}^2\right)}{(p_1+p_2)(1-p_0)^2}.
\end{align}

Similarly, the expressions of $\mathbb{E}\{D_j^2|G_m\}$ and $\mathbb{E}\{D_j^2|G_{m'}\}$ can be obtained as follows:
\begin{align}
&\quad\mathbb{E}\{D_{j}^2|G_m\}\\\notag
&=\sum_{k=1}^{\infty}\!(\!\frac{MT}{2}(2k\!-\!1))^2p_0^{k\!-\!1}p_2+ \sum_{k=1}^{\infty}\!(\!\frac{MT}{2}(2k))^2p_0^{k\!-\!1}(1\!-\!p_2)p_1\\\notag
&=\frac{M^2T^2[p_2(p_0^2+6p_0+1)+4p_1(1-p_2)(1+p_0)]}{4(1-p_0)^3},
\end{align}
and
\begin{align}
\mathbb{E}\{D_{j}^2|G_{m'}\}\!\!=\!\!\frac{M^2T^2[p_1(p_0^2\!+\!6p_0\!+\!1)\!+\!4p_2(1\!-\!p_1)(1\!+\!p_0)]}{4(1\!-\!p_0)^3}.
\end{align}
Therefore, $\mathbb{E}\{D_{j}^2\}$ can be expressed as:
\begin{align}\label{formula22}
&\mathbb{E}\{D_{j}^2\}\\\notag =&P(G_m)\mathbb{E}\{D_{j}^2|G_m\}+P(G_{m'})\mathbb{E}\{D_{j}^2|G_{m'}\}\\\notag
=&\frac{M^2T^2(2(1\!\!+\!\!p_0)(p_1^2(1\!\!-\!\!p_2)\!+\!{p_2}^2(1\!\!-\!\!p_1)))\!+\!p_1p_2(p_0^2\!\!+\!\!6p_0\!\!+\!\!1)}{2(p_1\!\!+\!\!p_2)(1\!\!-\!\!p_0)^3},
\end{align}
which completes the proof.

\section{Proof for Theorem $3$}
It is straightforward to show that $D_j$ and $S_{j-1}$ are independent of each other,which leads to the following:
\begin{align}\label{eq8}
\bar{\Delta}_m^{T\!D\!M\!A\!-\!R\!T}=\frac{\mathbb{E}\{Q_j\}}{\mathbb{E}\{D_j\}}=\mathbb{E}\{S_{j-1}\}+\frac{\mathbb{E}\{D_j^2\}/2}{\mathbb{E}\{D_j\}}.
\end{align}

\subsubsection{Evaluation of $\mathbb{E}\{S_{j-1}\}$}
as shown in  Fig.\ref{update process_exp} (c), the system time can be divided into two parts as follows:
\begin{align}\label{}
 S_{j-1}=\Delta t_1+\Delta t_2,
\end{align}
where $\Delta t_1$ is the waiting time of the transmitted update from its generation to the start of its first transmission,  and $\Delta t_2$ is the time duration of the transmitted update from the start of its first transmission to the end of its final transmission.

The evaluation of $S_{j-1}$ should be taken under the condition that the following two events occur, namely $\mathcal{C}_1$ and $\mathcal{C}_2$, where $\mathcal{C}_1$ denotes the event that there is at least one status update generated within the interval with duration $MT$ before the start of the first transmission time slot, and $\mathcal{C}_2$ denotes the event that the update is finally successfully transmitted. Thus, the expectation of $S_{j-1}$ can be expressed as follows:
\begin{align}
 \mathbb{E}\{S_{j-1}\}=\mathbb{E}\{\Delta t_1| \mathcal{C}\}+\mathbb{E}\{\Delta t_2| \mathcal{C}\},
\end{align}
where $\mathcal{C}=\mathcal{C}_1 \cap \mathcal{C}_2$.

By following the similar steps from (\ref{TDMA-NRT CDF}) to (\ref{waiting_time_TDMANRT}), the expression for $\mathbb{E}\{\Delta t_1| \mathcal{C}\}$ can be obtained as follows:
 \begin{align}
\mathbb{E}\{\Delta t_1|\mathcal{C}\}=\frac{MT}{1-e^{\lambda_mMT}}+\frac{1}{\lambda_m}.
 \end{align}

Note that the value of $\Delta t_2$ can be expressed as $\Delta t_2=kMT+T$, where $k$ is a random nonnegative integer. Then, by observing the fact that $\Delta t_2$ is independent of $\mathcal{C}_1$,  and $\mathcal{C}_1$ and $\mathcal{C}_2$ are independent of each other,  $\mathbb{E}\{\Delta t_2|\mathcal{C}\}$ can be expressed as:
\begin{align}\label{TDMA-RT Delta_t2}
\mathbb{E}\{\Delta t_2|\mathcal{C}\}\!\!=\!\!\sum_{k=0}^{\infty}(kMT+T)\frac{P(\Delta t_2=kMT+T,\mathcal{C}_2)}{P(\mathcal{C}_2)}.
\end{align}
It can be straightforwardly obtained that
\begin{align}\label{P(delta_t_C2)}
P(\Delta t_2\!=\!kMT\!+\!T,\mathcal{C}_2)\!=\!(e^{-\lambda_m MT})^k(1-P_{mm})^kP_{mm},
\end{align}
Hence, the expression for $P(\mathcal{C}_2)$ can be obtained as:
\begin{align}\label{PC2}
 P(\mathcal{C}_2)\!\!=\!\!\!\sum_{k=0}^{\infty}\!\!P(\!\Delta t_2\!\!=\!\!kMT\!\!+\!\!T,\mathcal{C}_2)\!\!=\!\!\frac{P_{mm}}{1\!\!-\!(1\!\!-\!P_{mm})e^{-\lambda_m\!MT}}.
\end{align}

By taking (\ref{P(delta_t_C2)}) and (\ref{PC2}) into (\ref{TDMA-RT Delta_t2}), the expression for $\mathbb{E}\{\Delta t_2|\mathcal{C}\}$ can be easily obtained:
\begin{align}\label{TDMA-RT expected value of waiting time for retransmission}
 \mathbb{E}\{\Delta t_2|\mathcal{C}\}=&\frac{MT(1-P_{mm})e^{-\lambda_mMT}}{1-(1-P_{mm})e^{-\lambda_mMT}}+T.
\end{align}
Thus,  $\mathbb{E}\{S_{j-1}\}$ can be expressed as:
\begin{align}\label{TDMA-RT System time expectation}
 \mathbb{E}\{S_{j-1}\}=\Gamma+\frac{MT(1-P_{mm})e^{-\lambda_mMT}}{1-(1-P_{mm})e^{-\lambda_mMT}}.
\end{align}
\subsubsection{Evaluation of $\mathbb{E}\{D_j\}$ and $\mathbb{E}\{D_j^2\}$}
\begin{figure}[!t]
  	\centering
	\vspace{0em}
	\setlength{\abovecaptionskip}{0em}   
	\setlength{\belowcaptionskip}{-2em}
  \includegraphics[width=1.5in]{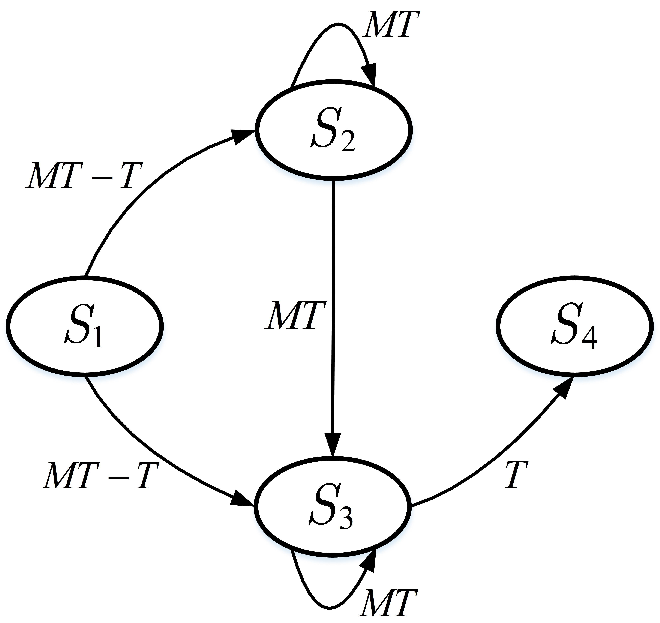}\\
  \caption{Illustration of the status updating process for TDMA-RT scheme by a Markov chain. The expressions along side the arrow lines denote the corresponding duration spent by the state transition.}\label{TDMA Markov}
\end{figure}
Fig. \ref{TDMA Markov} illustrates the state transition process from the end instant of the ($j-1$)-th successful transmission (state $S_1$) to the end instant of the $j$-th successful transmission (state $S_4$). Obviously, this process can be modeled as a Markov chain process with an absorbing wall, i.e., state $S_4$. Note that
$S_2$ and $S_3$ are transient states, which denote that there is no status update data to be transmitted and there exists status update data to be transmitted right at the beginning of a transmission slot, respectively. The  corresponding probability transition matrix is given by:
\begin{align}\label{}
\mathbf{\pi}&=
\begin{bmatrix}
 \mathbf{Q} & \mathbf{R} \\
\mathbf{0} & \mathbf{1}\\
\end{bmatrix},
\end{align}
where
\begin{align}\label{}
\small
\mathbf{Q}\!=\!\begin{bmatrix}
 0 & e^{-\lambda_mMT} & 1-e^{-\lambda_mMT}  \\
0 &  e^{-\lambda_mMT} & 1-e^{-\lambda_mMT} \\
0 & 0 & 1-P_{mm} \\
\end{bmatrix},
\mathbf{R}\!=\!\begin{bmatrix}
0  \\
0  \\
P_{mm} \\
\end{bmatrix}.
\end{align}

Note that $D_j$ is the total time spent from the start state $S_1$ to the absorbing state $S_4$,
which can be expressed as $D_j=(n-1)MT$, where $n$ is the number of transition steps from $S_1$ to $S_4$. Therefore, $\mathbb{E}\{D_j\}$ can be expressed by:
\begin{align}\label{TDMA-RT E(Dj)}
 \mathbb{E}\{D_j\}=& \mathbb{E}\{(n-1)MT\},
\end{align}
which indicates that it is essential to evaluate  $\mathbb{E}\{n\}$ to obtain the expression for $\mathbb{E}\{D_j\}$.

According to the absorbing Markov chain theory \cite{kemeny1976}, the expected number of steps from a transient state $S_i$ to the absorbing state can be obtained by a fundamental matrix $\mathbf{N}$, which is given by:
\begin{align}\label{matrix N}
\mathbf{N}=\sum_{k=0}^\infty \mathbf{Q}^k=(\mathbf{I}_3-\mathbf{Q})^{-1},
\end{align}
where $\mathbf{I}_3$ is a $3 \times 3$ identity matrix, and the $(i,j)$-th entry of $\mathbf{N}$ denotes the expected number of visits before being absorbed from the transient state $S_i$ to the transient state $S_j$.
It is noteworthy that the initial state is also counted for $\mathbf{N}$.
The expected number of steps from the transient state $S_i$ to the absorbing state can be obtained by the sum of the $i$-th row of the matrix $\mathbf{N}$, thus, $\mathbb{E}\{n\}$ can be expressed as:
\begin{align}\label{TDMA-RT E(n)}
\mathbb{E}\{n\}=&\mathbf{v}(1),
\end{align}
where
\begin{align}\label{TDMA-RT vector v}
\mathbf{v}=\mathbf{N}\mathbf{c}=
\begin{bmatrix}
\frac{1+P_{mm}-e^{-\lambda_mMT}}{(1-e^{-\lambda_mMT})P_{mm}}\\
\frac{1+P_{mm}-e^{-\lambda_mMT}}{(1-e^{-\lambda_mMT})P_{mm}} \\
\frac{1}{P_{mm}}
\end{bmatrix},
 \end{align}
$\mathbf{c}$ is a column vector with all elements $1$.

By taking (\ref{TDMA-RT E(n)}) into (\ref{TDMA-RT E(Dj)}), the expression for $\mathbb{E}\{D_j\}$ can be obtained, which can be expressed as:
\begin{align}\label{TDMA-RT expectation of the interval of each update}
 \mathbb{E}\{D_j\}=&\Psi.
\end{align}

Furthermore, $\mathbb{E}\{D_j^2\}$ can be expressed as:
 \begin{align}\label{TDMA-RT E(Dj^2)}
 \mathbb{E}\{D_j^2\}=& \mathbb{E}\{[(n-1)MT]^2\}\\\notag
=&M^2T^2(\sigma^2(n)+ \mathbb{E}^2\{n\}-2 \mathbb{E}\{n\}+1),
\end{align}
where $\sigma^2(n)$ is the variance of $n$.

Based on the  fundamental matrix $\mathbf{N}$ and vector $\mathbf{v}$, the variance of the number of steps before being absorbed when starting from transient state $S_i$ can be obtained by the $i$-th entry of the vector $\mathbf{\varphi}$, which is given by:
\begin{align}\label{vector}
\mathbf{\varphi}=(2\mathbf{\mathbf{N}}-\mathbf{\mathbf{I}}_3)\mathbf{v}-\mathbf{v}_{sq}
=
\begin{bmatrix}
s_1\\
s_2 \\
s_3
\end{bmatrix},
\end{align}
where $\mathbf{v}_{sq}$ is the Hadamard product of $\mathbf{v}$ with itself, which can be expressed as:
\begin{align}\label{}
&\mathbf{v}_{sq}=
\begin{bmatrix}
(\frac{1+P_{mm}-e^{-\lambda_mMT}}{(1-e^{-\lambda_mMT})P_{mm}})^2\\
(\frac{1+P_{mm}-e^{-\lambda_mMT}}{(1-e^{-\lambda_mMT})P_{mm}})^2 \\
(\frac{1}{P_{mm}})^2
\end{bmatrix},
\end{align}
and
\begin{align}\label{}
&s_1\!\!=\!\!\frac{(1\!+\!P_{mm}\!-\!e^{-\lambda_mMT})(e^{-\lambda_mMT}P_{mm}\!\!+\!e^{-\lambda_mMT}\!\!-\!1)}{(1\!-\!e^{-\lambda_mMT})^2P_{mm}^2}\!\!+\!\!\frac{2}{P_{mm}^2}, \\ &s_2\!\!=\!\!\frac{(1\!+\!P_{mm}\!-\!e^{-\lambda_mMT})(e^{-\lambda_mMT}P_{mm}\!+\!e^{-\lambda_mMT}\!\!-\!\!1)}{(1\!-\!e^{-\lambda_mMT})^2P_{mm}^2}\!\!+\!\!\frac{2}{P_{mm}^2},\\
&s_3=\frac{1-P_{mm}}{P_{mm}^2}.
\end{align}
Then, $\sigma^2(n)$ can be expressed as:
\begin{align}\label{TDMA-RT variance}
\sigma^2(n)=\mathbf{\varphi}(1)=s_1.
\end{align}

By taking (\ref{TDMA-RT E(n)}) and (\ref{TDMA-RT variance}) into (\ref{TDMA-RT E(Dj^2)}), the expression for $\mathbb{E}\{D_j^2\}$ can be obtained, which can be expressed as:
\begin{align}\label{TDMA-RT expectation of time squared per update interval}
&\mathbb{E}\{D_j^2\}=\Lambda.
\end{align}
And the proof is complete.

\section{Proof for Lemma $1$}
\begin{figure}[!t]
  \centering
  	\setlength{\abovecaptionskip}{0em}   
	\setlength{\belowcaptionskip}{0em}
  \includegraphics[width=1.4in]{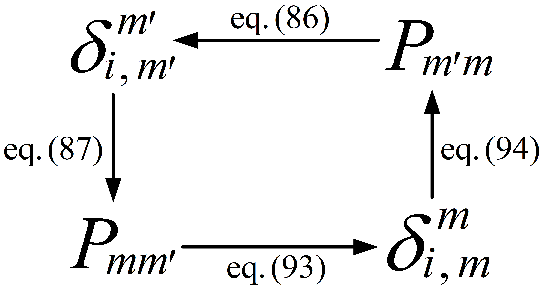}\\
  \caption{Relationships of $\delta_{i,m'}^{m'}$, $\delta_{i,m}^{m}$, $P_{mm'}$ and $P_{m'm}$ in NOMA-RT scheme.}\label{delta_PP}
\end{figure}
\begin{figure}[!t]
  	\centering
	\vspace{0em}
	\setlength{\abovecaptionskip}{0em}   
	\setlength{\belowcaptionskip}{-4em}
  \includegraphics[width=1.5in]{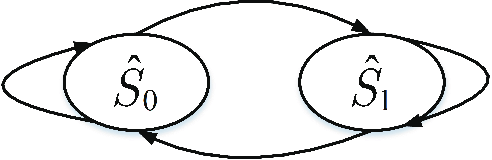}\\
  \caption{State transition diagram for $\delta_{i,m'}^{m'}$ in consecutive frames in NOMA-RT scheme.}\label{p4p5 Markov}
\end{figure}
Note that, $P_{mm'}$ and $P_{m'm}$ and the distributions of $\delta_{i,m'}^{m'}$ and $\delta_{i,m}^{m}$ in the steady state are coupled, as shown in Fig. \ref{delta_PP}. Thus, the key to evaluate $P_{mm'}$ is to establish equations for its relationships with
$P_{m'm}$ and the distributions of $\delta_{i,m'}^{m'}$ and $\delta_{i,m}^{m}$,
and then solve them.

Given $P_{m'm}$, the distribution of $\delta_{i,m'}^{m'}$ in steady state can be obtained as follows.
The transition process of $\delta_{i,m'}^{m'}$ for consecutive frames can be modeled as a Markov chain as shown in Fig. \ref{p4p5 Markov}.  $\hat{S}_0$ and $\hat{S}_1$ denote $\delta_{i,m'}^{m'}=0$ and $\delta_{i,m'}^{m'}=1$, respectively. The corresponding transition matrix can be expressed as:
\begin{align}\label{NOMA Markov}
\mathbf{\hat{\pi}}=
\begin{bmatrix}
 \hat{\pi}_{00} & \hat{\pi}_{01} \\
 \hat{\pi}_{10} & \hat{\pi}_{11}
\end{bmatrix},
\end{align}
where
\begin{align}\label{}
\hat{\pi}_{00}=&(1-e^{-\frac{\lambda_{m'}MT}{2}})e^{-\frac{\lambda_{m'}MT}{2}}P_{m'm}+e^{-\lambda_{m'}MT},\\
\hat{\pi}_{01}=&a_1P_{m'm}+(1-e^{-\lambda_{m'}MT}),\\
\hat{\pi}_{10}=&b_1P_{m'm}+P_{m'm'}e^{-\lambda_{m'}MT},\\
\hat{\pi}_{11}=&(1-P_{m'm'})(1-e^{-\frac{\lambda_{m'}MT}{2}}P_{m'm})+\\\notag
&P_{m'm'}(e^{-\frac{\lambda_{m'}MT}{2}}(e^{-\frac{\lambda_{m'}MT}{2}}\!-\!1)P_{m'm}\!+\!1\!-\!e^{-\lambda_{m'}MT}).
\end{align}

Denote the steady state probabilities for $\delta_{i,m'}^{m'}=0$ and $\delta_{i,m'}^{m'}=1$ by $\hat{\theta}_0$ and $\hat{\theta}_1$, respectively. Then, we have:
\begin{align}\label{delta m'm'}
\begin{bmatrix}
\hat{\theta}_0 & \hat{\theta}_1
\end{bmatrix}
\begin{bmatrix}
\hat{\pi}_{00} & \hat{\pi}_{01} \\
 \hat{\pi}_{10} & \hat{\pi}_{11}
\end{bmatrix}
=
\begin{bmatrix}
\hat{\theta}_0 & \hat{\theta}_1
\end{bmatrix}.
\end{align}

Given $\hat{\theta}_0$ and $\hat{\theta}_1$, $P_{mm'}$ can be expressed as:
\begin{align}\label{P_{mm}m'-1}
P_{mm'}=&\hat{\theta}_1P(R_{i,m'}^{m}\!\!>\!\!\frac{N}{T}|\delta_{i,m'}^{m'}\!\!=\!\!1)\!+\!\hat{\theta}_0P(R_{i,m'}^m\!\!>\!\!\frac{N}{T}|\delta_{i,m'}^{m'}\!\!=\!\!0)\notag\\
=&\hat{\theta}_1\Theta+\hat{\theta}_0e^{-\frac{\varepsilon}{P^s}}.
\end{align}

Similarly, the transition process of $\delta_{i,m}^{m}$ for consecutive frames can also be modeled as a Markov process, with the following transition matrix:
\begin{align}\label{NOMA Markov}
\mathbf{\tilde{\pi}}=
\begin{bmatrix}
\tilde{\pi}_{00} & \tilde{\pi}_{01} \\
\tilde{\pi}_{10} & \tilde{\pi}_{11}
\end{bmatrix}.
\end{align}
where
\begin{align}\label{}
\tilde{\pi}_{00}=&(1-e^{-\frac{\lambda_mMT}{2}})e^{-\frac{\lambda_mMT}{2}}P_{mm'}+e^{-\lambda_mMT},\\
\tilde{\pi}_{01}=&a_2P_{mm'}+(1-e^{-\lambda_mMT}),\\
\tilde{\pi}_{10}=&b_2P_{mm'}+P_{mm}e^{-\lambda_mMT},\\
\tilde{\pi}_{11}=&(1-P_{mm})(1-e^{-\frac{\lambda_mMT}{2}}P_{mm'})+\\\notag
&P_{mm}(e^{-\frac{\lambda_mMT}{2}}(e^{-\frac{\lambda_mMT}{2}}\!-\!1)P_{mm'}\!+\!1\!-\!e^{-\lambda_mMT}),
\end{align}

Denoted the steady state probabilities  for $\delta_{i,m}^{m}=0$ and $\delta_{i,m}^{m}=1$ by $\tilde{\theta}_0$ and $\tilde{\theta}_1$, respectively, which leads to the following:
\begin{align}\label{delta mm}
\begin{bmatrix}
\tilde{\theta}_0 & \tilde{\theta}_1
\end{bmatrix}
\begin{bmatrix}
\tilde{\pi}_{00} & \tilde{\pi}_{01} \\
\tilde{\pi}_{10} & \tilde{\pi}_{11}
\end{bmatrix}
=
\begin{bmatrix}
\tilde{\theta}_0 & \tilde{\theta}_1
\end{bmatrix}.
\end{align}

Given $\tilde{\theta}_0$ and $\tilde{\theta}_1$, $P_{m'm}$ can be expressed as:
\begin{align}\label{P_{mm}'m-1}
P_{m'm}=&\tilde{\theta}_1P(R_{i,m}^{m'}\!\!>\!\!\frac{N}{T}|\delta_{i,m}^{m}\!\!=\!\!1)\!+\!\tilde{\theta}_0P(R_{i,m}^{m'}\!\!>\!\!\frac{N}{T}|\delta_{i,m}^{m}\!\!=\!\!0)\notag\\
=&\tilde{\theta}_1\Theta\!+\tilde{\theta}_0e^{-\frac{\varepsilon}{P^s}},
\end{align}

By combining (\ref{delta m'm'}), (\ref{P_{mm}m'-1}), (\ref{delta mm}) and (\ref{P_{mm}'m-1}), the expression for $P_{mm'}$ can be obtained, and the proof is complete.

\section{Proof for Theorem $4$}
$\mathbb{E}\{D_jS_{j-1}\}$ can be written as follows:
\begin{align}\label{E(DS)-1}
\mathbb{E}\{D_jS_{j-1}\}
&\!\!=\!\!P\!(Q_m\!)\mathbb{E}\!\{D_jS_{j\!-\!1}|Q_m\!\}\!+\!P\!(Q_{m'}\!)\mathbb{E}\!\{D_jS_{j\!-\!1}|Q_{m'}\!\}\notag\\
&\overset{(a)}{=}P(Q_m)\mathbb{E}\{D_j|Q_m\}\mathbb{E}\{S_{j-1}|Q_m\}\notag\\
&\quad+P(Q_{m'})\mathbb{E}\{D_j|Q_{m'}\}\mathbb{E}\{S_{j-1}|Q_{m'}\},
\end{align}
where $Q_m$ (or $Q_{m'}$) denotes the event that the transmission of the $(j-1)$-th successfully delivered update is finished at the end of the $m$-th (or $m'$)-th time slot. Note that step (a) follows from the fact that, given $Q_m$ (or $Q_{m'}$), $S_{j-1}$ and $D_j$ are independent of each other.

\begin{figure}[!t]
  \centering
  	\setlength{\abovecaptionskip}{0em}   
	\setlength{\belowcaptionskip}{-2em}
  \includegraphics[width=1.2in]{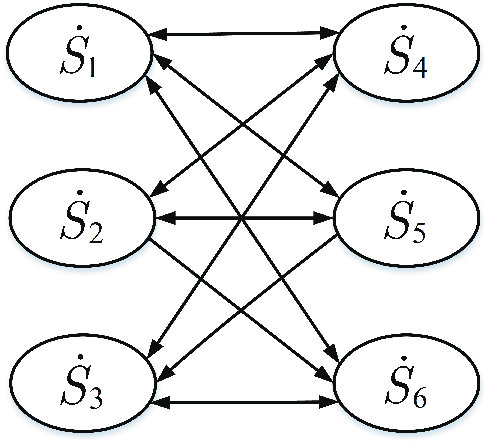}\\
  \caption{State transition diagram for the states at the end of the $m$-th and $m'$-th slot of each frame in NOMA-RT scheme.}\label{RTRT Markov}
\end{figure}
\begin{figure*}[!t]
\small
\vspace{0em}
\begin{align}\label{dot_pi}
\dot{\mathbf{\pi}}=
\begin{bmatrix}\!
 0 & 0 & 0  & (1\!\!-\!e^{-\frac{\lambda_mMT}{2}})P_{mm'} & e^{-\frac{\lambda_mMT}{2}} & (1\!\!-\!e^{-\frac{\lambda_mMT}{2}})(1\!\!-\!\!P_{mm'}) \\
0 & 0 & 0 & (1\!\!-\!e^{-\frac{\lambda_mMT}{2}})P_{mm'} & e^{-\frac{\lambda_mMT}{2}} & (1\!\!-\!e^{-\frac{\lambda_mMT}{2}})(1\!\!-\!\!P_{mm'})\\
0 & 0 & 0 & P_{mm'} & 0 & (1\!-\!P_{mm'}) \\
(1\!\!-\!e^{-\frac{\lambda_mMT}{2}})P_{mm} & e^{-\frac{\lambda_mMT}{2}} & (1\!\!-\!e^{-\frac{\lambda_mMT}{2}})(1\!\!-\!\!P_{mm}) & 0 & 0 & 0\\
(1\!\!-\!e^{-\frac{\lambda_mMT}{2}})P_{mm} & e^{-\frac{\lambda_mMT}{2}} & (1\!\!-\!e^{-\frac{\lambda_mMT}{2}})(1\!\!-\!\!P_{mm}) & 0 &0 & 0\\
P_{mm} & 0 & (1\!\!-\!\!P_{mm}) & 0 & 0 & 0\\
\end{bmatrix}.
\end{align}
\vspace{-2em}
\end{figure*}
To obtain $P(Q_m)$ and $P(Q_{m'})$, it is necessary to consider all possible states (from the receiver perspective) at the end of $m$-th and $m'$-th slot of each frame, whose transitions can be modeled as a Markov chain, as shown in
Fig. \ref{RTRT Markov}. In Fig. \ref{RTRT Markov}, $\dot{S}_1$ (or $\dot{S}_4$) denotes the state that
a new status update packet arrives at the receiver successfully within the $m$-th slot (or $m'$-th slot).
$\dot{S}_2$ (or $\dot{S}_5$) denotes the state that there is no new status update data received by the receiver within  the $m$-th slot (or $m'$-th slot), due to the reason that there's no status data to be transmitted within the time slot. $\dot{S}_3$ (or $\dot{S}_6$) also denotes the state that there is no new status update data received by the receiver within  the $m$-th slot (or $m'$-th slot), due to the transmission failure.
The corresponding probability transition matrix $\mathbf{\dot{\pi}}$ can be expressed as shown in (\ref{dot_pi}) at the top of next page.

Denote the steady state probability for $\dot{S}_i$ by $\gamma_i$, $i=1, \cdots, 6$.
The expression of $\gamma_i$ can be obtained by solving the following steady state equation:
\begin{align}\label{gamma}
\setlength{\arraycolsep}{2pt}
\begin{bmatrix}
\gamma_1 & \gamma_2 & \gamma_3 & \gamma_4 & \gamma_5 & \gamma_6
\end{bmatrix}
\mathbf{\dot{\pi}}
=
\setlength{\arraycolsep}{2pt}
\begin{bmatrix}
\gamma_1 & \gamma_2 & \gamma_3 & \gamma_4 & \gamma_5 & \gamma_6
\end{bmatrix}.
\end{align}

Particularly, $\gamma_1$ and $\gamma_4$ can be expressed by (\ref{eq:gamma1}) and (\ref{eq:gamma2}), respectively.

Then, according to the definitions of $P(Q_m)$ and $P(Q_{m'})$, it can be easily obtained that:
\begin{align}\label{Qm}
P(Q_m)=\frac{\gamma_1}{\gamma_1+\gamma_4},
P(Q_{m'})=\frac{\gamma_4}{\gamma_1+\gamma_4}.
\end{align}

The next task is to evaluate $\mathbb{E}\{S_{j-1}|Q_m\}$, where $\mathbb{E}\{S_{j-1}|Q_{m'}\}$ can be obtained similarly.
As shown in  Fig.\ref{update process_exp} (d), $S_{j-1}$ can be divided into two parts as follows:
\begin{align}\label{}
 S_{j-1}=\Delta t_1+\Delta t_2,
\end{align}
where $\Delta t_1$ is the waiting time of the transmitted update from its generation to the start of its first transmission,  and $\Delta t_2$ is the time duration of the transmitted update from the start of its first transmission to the end of its final transmission.

\begin{figure}[!t]
  	\centering
	\vspace{0em}
	\setlength{\abovecaptionskip}{0em}   
	\setlength{\belowcaptionskip}{-2em}
  \includegraphics[width=2.0in]{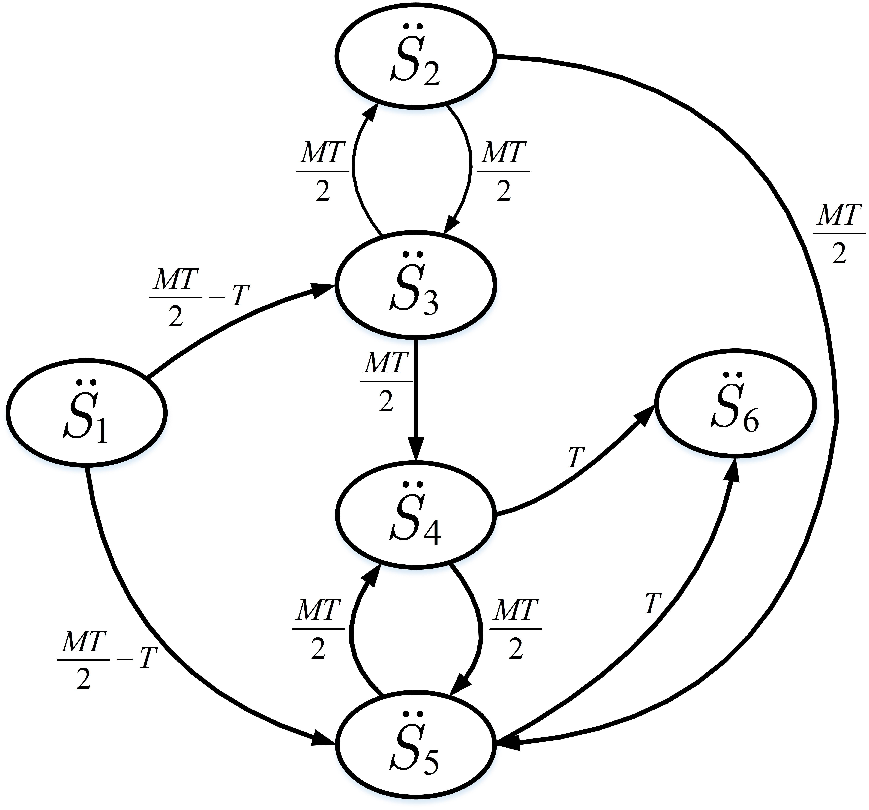}\\
  \caption{Illustration of the state transition process from the end instant of the ($j-1$)-th successful transmission to the end instant of the $j$-th successful transmission. The expressions along side the arrow lines denote the corresponding duration spent by the state transition.}\label{NOMA Markov}
\end{figure}
Note that the evaluation of $S_{j-1}$ should be taken under the condition that the following two events occur, namely $\mathcal{D}_1$ and $\mathcal{D}_2$, where $\mathcal{D}_1$ denotes the event that there is  at least one status update generated within the interval with duration $\frac{MT}{2}$ before the start of the first transmission time slot, and $\mathcal{D}_2$ denotes the event that the status update is finally successfully transmitted within the transmitting time slot. It is noteworthy that $\mathcal{D}_2$ can be divided into
two disjoint events, i.e., $\mathcal{D}_2=\mathcal{D}_{2m} \cup \mathcal{D}_{2m'}$, where $\mathcal{D}_{2m}$ and $\mathcal{D}_{2m'}$ denote the transmission is completed within the $m$-th time slot and $m'$-th time slot of a frame, respectively. Thus, the expression of $\mathbb{E}\{S_{j-1}|Q_m\}$ can be written as follows:
\begin{align}
&\mathbb{E}\{S_{j-1}|Q_m\}\!\!=\!\!\mathbb{E}\{\Delta t_1|\mathcal{D}_1,\mathcal{D}_{2m}\}\!+\!\mathbb{E}\{\Delta t_2|\mathcal{D}_1,\mathcal{D}_{2m}\}.
\end{align}

By following the similar steps from (\ref{distribution}) to (\ref{NOMA-NRT E(Sj-1|Gm)}), the expression for $\mathbb{E}\{\Delta t_1|\mathcal{D}_1,\mathcal{D}_{2m}\}$ can be obtained as follows:
\begin{align}\label{NOMA E(Delta t)}
\mathbb{E}\{\Delta t_1|\mathcal{D}_1,\mathcal{D}_{2m}\}
 =&\bar{\Omega}-T.
 \end{align}

Rewrite $\Delta t_2$ as $\Delta t_2=\frac{kMT}{2}+T$,  where $k$ is a random nonnegative integer, and $\mathbb{E}\{\Delta t_2|\mathcal{D}_1,\mathcal{D}_{2m}\}$ can be expressed as follows:
\begin{align}\label{E_delta_t2}
\quad &\mathbb{E}\{\Delta t_2|\mathcal{D}_1,\mathcal{D}_{2m}\}\\\notag
&=\sum_{k=0}^{\infty}(\frac{kMT}{2}+T)\frac{P(\Delta t_2=(\frac{kMT}{2}+T) , D_1,D_{2m})}{P(D_1,D_{2m})}\\\notag
&=\sum_{k=0}^{\infty}(\frac{kMT}{2}+T)\frac{P(\Delta t_2=(\frac{kMT}{2}+T) ,D_{2m})}{P(D_{2m})}.
\end{align}
When k is an even number, we have:
\begin{align}\label{E_delta_t2_even}
P(\Delta t_2\!=\!(\frac{kMT}{2}\!+\!T), D_{2m})\!=\!P_{mm}(e^{-\frac{\lambda_mMT}{2}})^{k}P_{m0}^{\frac{k}{2}},
\end{align}
where $P_{m0}=(1-P_{mm})(1-P_{mm'})$, and when k is an odd number, we have:
\begin{align}\label{E_delta_t2_odd}
P(\!\Delta t_2\!=\!(\!\frac{kMT}{2}\!\!+\!\!T), D_{2m}\!)\!=\!P_{mm}(e^{-\frac{\lambda_mMT}{2}})^k(1\!\!-\!\!P_{mm'}\!)P_{m0}^{\frac{k\!-\!1}{2}}.
\end{align}
Then, the expression for $P(D_{2m})$ can be obtained, which is given by:
\begin{align}\label{P_D_2m}
P(D_{2m})&\!=\!\sum_{k=0}^{\infty}P(\Delta t_2=(\frac{kMT}{2}+T), D_{2m})\\\notag
&=\frac{P_{mm}(1+(1-P_{mm'})e^{-\frac{\lambda_mMT}{2}})}{1-P_{m0}e^{-\lambda_mMT}}.
\end{align}
By taking (\ref{E_delta_t2_even})-(\ref{P_D_2m}) into (\ref{E_delta_t2}), it can be obtained that:
\begin{align}\label{}
\quad& \mathbb{E}\{\Delta t_2|\mathcal{D}_1,\mathcal{D}_{2m}\}\\\notag
=&\sum_{k=0}^{\infty}\frac{(2k+1)\frac{MT}{2}P_{m0}^k(1-P_{mm'})P_{mm}(e^{-\frac{\lambda_mMT}{2}})^{2k+1}}{P(D_{2m})}\\\notag
&+\sum_{k=0}^{\infty}\frac{(2k)\frac{MT}{2}P_{m0}^kP_{mm}(e^{-\frac{\lambda_mMT}{2}})^{2k}}{P(D_{2m})}+T=Y+T.
\end{align}
Thus, the expression for $\mathbb{E}\{S_{j-1}|Q_m\}$ can be obtained as:
\begin{align}\label{E(Sm)}
\mathbb{E}\{S_{j-1}|Q_m\}=&\bar{\Omega}+Y.
\end{align}
Similarly, it can be obtained that:
\begin{align}\label{E(Sm')}
\mathbb{E}\{S_{j-1}|Q_{m'}\}=&\bar{\Omega}+Y'.
\end{align}

\begin{figure*}[!t]
\vspace{0em}
\begin{align}
L=&P_{mm'}(P_{mm}^2P_{mm'}+P_{mm}^2-5P_{mm}P_{mm'}+4P_{mm'}),
\quad S=P_{mm'}(P_{mm}^2P_{mm'}-P_{mm}^2-3P_{mm}P_{mm'}+4P_{mm'})\label{ls1}\\ R=&2P_{mm}^2P_{mm'}^2-7P_{mm}^2P_{mm'}+4P_{mm}^2+P_{mm}P_{mm'}^2+6P_{mm}P_{mm'}-4P_{mm'}^2\label{r1}\\ T=&4P_{m0}(1-e^{-\lambda_mMT})^2+(P_{mm'}+e^{-\frac{\lambda_mMT}{2}}(P_{m0}+P_{mm}-P_{mm'}-1)+e^{-\lambda_mMT}(P_{m0}-P_{mm}+1)-2)^2+\label{t1}\\\notag
&P_{mm}P_{mm'}(1-P_{mm'}+e^{-2\lambda_mMT}(1-P_{mm}))\\
L'=&P_{mm}(P_{mm'}^2P_{mm}+P_{mm'}^2-5P_{mm'}P_{mm}+4P_{mm}),
\quad S'=P_{mm}(P_{mm'}^2P_{mm}-P_{mm'}^2-3P_{mm'}P_{mm}+4P_{mm}\label{ls2}\\ R'=&2P_{mm'}^2P_{mm}^2-7P_{mm'}^2P_{mm}+4P_{mm'}^2+P_{mm'}P_{mm}^2+6P_{mm'}P_{mm}-4P_{mm}^2\label{r2}\\ T'=&4P_{m0}(1-e^{-\lambda_mMT})^2+(P_{mm}+e^{-\frac{\lambda_mMT}{2}}(P_{m0}+P_{mm'}-P_{mm}-1)+e^{-\lambda_mMT}(P_{m0}-P_{mm'}+1)-2)^2+\label{t2}\\\notag
&P_{mm'}P_{mm}(1-P_{mm}+e^{-2\lambda_mMT}(1-P_{mm'}))
\end{align}
\vspace{-3em}
\end{figure*}
The next step is to evaluate $\mathbb{E}\{D_j|Q_m\}$.
As shown in Fig. \ref{NOMA Markov}, given $Q_m$, the state transition process from the end instant of the ($j-1$)-th successful transmission (state $\ddot{S}_1$) to the end instant of the $j$-th successful transmission (state $\ddot{S}_6$) can be modeled as a Markov process with an absorbing wall,  where $\ddot{S}_1$, $\ddot{S}_2$, $\ddot{S}_3$, $\ddot{S}_4$, and $\ddot{S}_5$ are the transient states, and $\ddot{S}_6$ is the absorbing state.
State $\ddot{S}_2$ (or $\ddot{S}_3$) denotes the state that there is no one status update packet to be transmitted within the $m$-th (or $m'$-th) slot, and state $\ddot{S}_4$ (or $\ddot{S}_5$) denotes the state that there is status update packet to be transmitted within the $m$-th (or $m'$)-th slot. The probability transition matrix $\mathbf{\ddot{\mathbf{\pi}}}$ for the absorbing Markov chain is given by:
\begin{align}\label{}
\mathbf{\ddot{\mathbf{\pi}}}
&=
\begin{bmatrix}
 \mathbf{Q} & \mathbf{R} \\
\mathbf{0} & \mathbf{1}\\
\end{bmatrix},
\end{align}
where
\begin{align}\label{}
\small
\setlength{\arraycolsep}{0.7pt}
\mathbf{Q}\!\!=\!\!
\begin{bmatrix}
 0 & 0 & e^{\!-\frac{\lambda_m\!MT}{2}}  & 0 & 1\!-\!e^{\!-\frac{\lambda_m\!MT}{2}}  \\
0 & 0 & e^{\!-\frac{\lambda_m\!MT}{2}} & 0 & 1\!\!-\!e^{\!-\frac{\lambda_m\!MT}{2}} \\
0 & e^{\!-\frac{\lambda_m\!MT}{2}} & 0 & 1\!\!-\!e^{\!-\frac{\lambda_m\!MT}{2}} & 0  \\
0 & 0 & 0 & 0 & 1\!\!-\!P_{mm} \\
0 & 0 & 0 & 1\!\!-\!P_{mm'} &0
\end{bmatrix},
\mathbf{R}\!\!=\!\!
\begin{bmatrix}
0 \\
0 \\
0\\
P_{mm}\\
P_{mm'}
\end{bmatrix}.
\end{align}

It can be observed that $D_j$ is the total time elapsed from  state $\ddot{S}_1$ to state $\ddot{S}_6$, which
can be expressed as $D_j=(n-1)\frac{MT}{2}$, where $n$ is the number of steps from $\ddot{S}_1$ to $\ddot{S}_6$. Hence, it can be obtained that:
\begin{align}\label{NOMA-RT E(Dj)}
 \mathbb{E}\{D_j|Q_m\}=& \mathbb{E}\{(n-1)\frac{MT}{2}\}.
\end{align}

By following the similar steps from (\ref{matrix N}) to (\ref{TDMA-RT E(n)}), the expression for $\mathbb{E}\{D_j|Q_m\}$ can be obtained, as shown in the following:
\begin{align}\label{E(Dm)}
&\quad \mathbb{E}\{D_j|Q_m\}=W.
\end{align}

By using the same method for $\mathbb{E}\{D_j|Q_{m}\}$, the expression for $\mathbb{E}\{D_j|Q_{m'}\}$ can be obtained as follows:
\begin{align}\label{E(Dm')}
&\quad \mathbb{E}\{D_j|Q_{m'}\}=W'.
\end{align}

By taking (\ref{Qm}), (\ref{E(Sm)}), (\ref{E(Sm')}), (\ref{E(Dm)}), and (\ref{E(Dm')}) into (\ref{E(DS)-1}), the expression for $\mathbb{E}\{D_jS_{j-1}\}$ can be straightforwardly obtained:
\begin{align}\label{}
 \mathbb{E}\{D_jS_{j-1}\}=\frac{\gamma_1(\bar{\Omega}+Y)W+\gamma_4(\bar{\Omega}+Y')W'}{(\gamma_1+\gamma_4)}.
\end{align}

Furthermore, the expression for $\mathbb{E}\{D_j\}$ can be obtained as follows:
\begin{align}\label{}
\mathbb{E}\{D_j\}=&\frac{\gamma_1W+\gamma_4W'}{\gamma_1+\gamma_4}.
\end{align}

By following the similar steps from (\ref{TDMA-RT E(Dj^2)}) to (\ref{TDMA-RT variance}), the expressions for $ \mathbb{E}\{D_j^2|Q_{m}\}$ and $\mathbb{E}\{D_j^2|Q_{m'}\}$ can be obtained, which are given by:
\begin{align}\label{}
&\quad \mathbb{E}\{D_j^2|Q_m\}\\\notag
&=\frac{M^2T^2}{4}(\frac{e^{-\frac{3\lambda_mMT}{2}}L+e^{-\lambda_mMT}R+e^{-\frac{\lambda_mMT}{2}}S+T}{(1-e^{-\lambda_mMT})^2(1-P_{m0})^2}),
\end{align}
where $L$, $S$, $R$, $T$ are shown in (\ref{ls1}), (\ref{r1}) and (\ref{t1}), and
\begin{align}\label{}
&\quad \mathbb{E}\{D_j^2|Q_{m'}\}\\\notag
&=\frac{M^2T^2}{4}(\frac{e^{-\frac{3\lambda_mMT}{2}}L'+e^{-\lambda_mMT}R'+e^{-\frac{\lambda_mMT}{2}}S'+T'}{(1-e^{-\lambda_mMT})^2(1-P_{m0})^2}),
\end{align}
where $L'$, $S'$, $R'$, $T'$ are shown in (\ref{ls2}), (\ref{r2}) and (\ref{t2}).

Thus, the expression for $\mathbb{E}\{D_j^2\}$ can be obtained as follows:
\begin{align}
\mathbb{E}\{D_j^2\}
\!\!=\!\!P(\!Q_m\!)\mathbb{E}\{D_j^2|Q_m\}\!\!+\!\!P(\!Q_{m'}\!)\mathbb{E}\{D_j^2|Q_{m'}\}\!\!=\!\!H.
\end{align}

The proof is complete.
\bibliographystyle{IEEEtran}
\bibliography{IEEEabrv,ref}
\end{document}